\pdfoutput=1
\documentclass[twocolumn]{aastex6}

\usepackage{amsmath}
\usepackage{graphicx}
\usepackage{lineno}
\usepackage{units}

\usepackage{hyperref}
\usepackage{aas_macros}

\newcommand{\Gp}[1] {\ensuremath{\left(#1\right)}}
\newcommand{\Gb}[1] {\ensuremath{\left[#1\right]}}

\begin{document}

\title{Search for astrophysical sources of neutrinos using cascade events in
IceCube}
\slugcomment{Submitted to The Astrophysical Journal}
\shortauthors{M.~G.~Aartsen et al.}

%\input{authors.tex}
% authors.tex

%\AuthorCallLimit=1
\fullcollaborationName{IceCube}

\author{
IceCube Collaboration:
M.~G.~Aartsen\altaffilmark{1},
M.~Ackermann\altaffilmark{2},
J.~Adams\altaffilmark{3},
J.~A.~Aguilar\altaffilmark{4},
M.~Ahlers\altaffilmark{5},
M.~Ahrens\altaffilmark{6},
I.~Al~Samarai\altaffilmark{7},
D.~Altmann\altaffilmark{8},
K.~Andeen\altaffilmark{9},
T.~Anderson\altaffilmark{10}
% ...
}
\altaffiltext{1}{Department of Physics, University of Adelaide, Adelaide, 5005, Australia}
\altaffiltext{2}{DESY, D-15735 Zeuthen, Germany}
\altaffiltext{3}{Dept.~of Physics and Astronomy, University of Canterbury, Private Bag 4800, Christchurch, New Zealand}
\altaffiltext{4}{Universit\'e Libre de Bruxelles, Science Faculty CP230, B-1050 Brussels, Belgium}
\altaffiltext{5}{Niels Bohr Institute, University of Copenhagen, DK-2100 Copenhagen, Denmark}
\altaffiltext{6}{Oskar Klein Centre and Dept.~of Physics, Stockholm University, SE-10691 Stockholm, Sweden}
\altaffiltext{7}{D\'epartement de physique nucl\'eaire et corpusculaire, Universit\'e de Gen\`eve, CH-1211 Gen\`eve, Switzerland}
\altaffiltext{8}{Erlangen Centre for Astroparticle Physics, Friedrich-Alexander-Universit\"at Erlangen-N\"urnberg, D-91058 Erlangen, Germany}
\altaffiltext{9}{Department of Physics, Marquette University, Milwaukee, WI, 53201, USA}
\altaffiltext{10}{Dept.~of Physics, Pennsylvania State University, University Park, PA 16802, USA}
% ...

% END: authors.tex

\received{July 28, 2017}

\begin{abstract}
  The IceCube neutrino observatory has established the existence of a flux
  of high-energy astrophysical neutrinos inconsistent with the expectation
  from atmospheric backgrounds at a significance greater than $5\sigma$.
  This flux has been observed in analyses of both track events from muon
  neutrino interactions and cascade events from interactions of all neutrino
  flavors.  Searches for astrophysical neutrino sources have focused on
  track events due to the significantly better angular resolution of track
  reconstructions.  To date, no such sources have been confirmed.  Here we
  present the first search for astrophysical neutrino sources using cascades
  interacting in IceCube with deposited energies as small as \unit[1]{TeV}.
  No significant clustering was observed in a selection of 263 cascades
  collected from May 2010 to May 2012.  We show that compared to the classic
  approach using tracks, this statistically-independent search offers
  improved sensitivity to sources in the southern sky, especially if the
  emission is spatially extended or follows a soft energy spectrum.  This
  enhancement is due to the low background from atmospheric neutrinos
  forming cascade events and the additional veto of atmospheric neutrinos at
  declinations $\lesssim-30^\circ$.
\end{abstract}

\maketitle

\cleardoublepage

\section{Introduction}

Neutrinos are promising messenger particles for astrophysical observations
due to their extremely small interaction cross-sections and lack of electric
charge.  They can travel enormous distances largely unimpeded by intervening
matter and undeflected by magnetic fields.  These properties make it
possible to associate neutrinos from distant sources with each other and
with known sources of electromagnetic radiation.  Furthermore, because
neutrinos are produced in high-energy hadronic interactions, observations of
astrophysical neutrinos will shed light on the still-elusive origins of the
highest-energy cosmic rays
\citep{Gaisser:Halzen:Stanev,Learned:Mannheim,Becker:2007sv}.

IceCube is the first $\unit{km^3}$-scale neutrino detector
\citep{Achterberg:2006md}.  Using an array of photomultiplier tubes (PMTs)
deployed deep in the antarctic glacial ice near the South Pole, it can
detect neutrinos of all flavors by collecting the Cherenkov light emitted by
the relativistic charged particles produced when neutrinos interact with
atomic nuclei in the ice.  Neutrinos produce one of two topologically
distinct signatures: \emph{tracks} and \emph{cascades}.  Charged current
(CC) muon neutrino interactions yield long-lived muons that can travel
several kilometers through the ice \citep{Chirkin:2004hz}, producing an
elongated track signature in the detector.  Charged current interactions of
other neutrino flavors, and all neutral current (NC) interactions, yield
hadronic and electromagnetic showers that typically range less than
$\unit[20]{m}$ \citep{Aartsen:2013vja}, with 90\% of the light emitted
within $\unit[4]{m}$ of the shower maximum~\citep{Radel:2012ij} | a short
distance compared to the scattering and absorption lengths of light in the
ice \citep{Aartsen:2013rt} as well as the spacing of the PMTs.  These
showers produce a nearly spherically symmetric cascade signature in light.

A flux of astrophysical neutrinos above $\sim\unit[60]{TeV}$ inconsistent
with the expectation from atmospheric backgrounds at greater than $5\sigma$
was first established using neutrinos interacting within the instrumented
volume of IceCube \citep{Aartsen:2014gkd,hese4}.  The majority of events
contributing to this measurement were cascades.  More recently, this flux
has been confirmed in an analysis of tracks from muon neutrinos above
$\sim\unit[300]{TeV}$ originating in the northern sky
\citep{Aartsen:2015rwa,Aartsen:2016xlq}.  No significant anisotropy has yet
been observed, and the neutrino flavor ratio at Earth is consistent with
1:1:1 \citep{Aartsen:2015ivb}.

Searches for astrophysical neutrino sources have traditionally focused on
track events because the elongated signature gives much better angular
resolution than can be obtained for cascades.  While ANTARES recently
reported the addition of a cascade selection to their all-sky search for
sources of steady neutrino emission \citep{TheANTARES:2015pba}, IceCube has
so far excluded cascades from its all-sky search \citep{Aartsen:2016oji}
except in the simplified analysis applied only to very high-energy contained
events \citep{hese4}.  

In this paper, we present the first all-sky search for astrophysical
neutrino sources producing cascades in IceCube with deposited energies as
small as \unit[1]{TeV}.  This analysis includes 263 cascades observed from
May 2010 to May 2012.  We find that, due to the relatively low rate of
atmospheric backgrounds in this sample, this search reduces the energy
threshold in the southern sky relative to previous IceCube work with tracks.
The sensitivity of this search is much less dependent on the declination,
spatial extension, and emission spectrum of a possible source.  In the
following sections, we begin with an overview of the detector, experimental
dataset, and statistical methods used in this analysis before reporting
results from the two-year sample and discussing directions for future work.

\section{IceCube}
\label{sec:icecube}

The IceCube detector (described in detail in
\citet{DETPAPER-1748-0221-12-03-P03012}) consists of 5160 Digital Optical
Modules (DOMs) buried in the glacial ice near the South Pole.  The DOMs are
mounted on 86 vertical ``strings'', with 60 DOMs on each string.  Each
string is connected to a central lab on the surface by a cable that provides
power and communication with the data acquisition (DAQ) system
\citep{Abbasi:2008aa}.  Seventy-eight of the strings are arranged in a
hexagonal grid with a spacing of \unit[125]{m}; on these strings, DOMs are
distributed uniformly from \unit[1450]{m} to \unit[2450]{m} below the
surface of the ice.  The remaining 8 strings make up the denser DeepCore
in-fill array \citep{Collaboration:2011ym}, with inter-string spacing of 30
to \unit[60]{m}.  The in-fill strings include 50 DOMs in the particularly
clear ice at depths of \unit[2100]{m} to \unit[2450]{m} and an additional 10
DOMs evenly spaced at depths of \unit[1750]{m} to \unit[1850]{m}.
Construction was performed during Austral summers starting in 2004.  A
nearly complete 79-string configuration began taking data in May 2010, and
the first year of data from the complete 86-string detector was taken from
May 2011 to May 2012.

Each DOM includes a \unit[25]{cm} diameter PMT \citep{Abbasi:2010vc} and
supporting electronics.   A local coincidence condition occurs when a DOM
and one of its nearest neighbors exceed a threshold of 1/4 of the mean
expected voltage from a single photoelectron (PE).  When at least eight DOMs
observe local coincidence within $\unit[6.4]{\mu s}$, the DAQ produces an
event consisting of \unit[400]{ns} digitized waveforms from all DOMs
observing local coincidence and \unit[75]{ns} waveforms from all other DOMs
exceeding the threshold.  The waveforms are then decomposed into series of
pulse arrival times and PE counts which are used to reconstruct the
trajectory and deposited energy of the relativistic particles in the
detector \citep{Ahrens:2003fg,Aartsen:2013vja}.

The simple eight-DOM trigger accepts neutrino-induced events with very high
efficiency.  Unfortunately, even deep in the glacial ice, cosmic ray-induced
atmospheric muons trigger the detector at an average rate of
\unit[2.7]{kHz}, overwhelming the trigger rate of atmospheric neutrinos
(\mbox{$\sim\unit[20]{mHz}$}) and rare astrophysical neutrinos.  An initial
data reduction step performed at the South Pole reduces the event rate by a
factor of 100 by rejecting lower-energy events that are consistent with
downgoing tracks.  The remaining dataset, still dominated by downgoing
muons, is transmitted to the northern hemisphere via satellite for further
analysis.

\section{Neutrino Selection}
\label{sec:neutrinos}

IceCube searches for moderate to high energy neutrinos generally exploit one
of two methods to reject the atmospheric muon background.  The largest
effective volume and best angular resolution are available when incoming
muon tracks are accepted.  This approach offers good performance for muon
neutrinos from the northern celestial hemisphere because only neutrinos can
survive passage through the intervening earth before producing upgoing muons
in the ice.  However, neutrino- and cosmic ray-induced downgoing muon tracks
entering the detector from above produce nearly indistinguishable event
topologies.  Astrophysical neutrinos from the southern sky can be identified
on a statistical basis if the neutrino spectrum is harder than the
atmospheric muon spectrum, but this strategy increases the energy threshold
to $\sim\unit[100]{TeV}$ in the southern sky, compared to only
$\sim\unit[1]{TeV}$ in the northern sky \citep{Aartsen:2014cva}.

An alternative method is to restrict the analysis to ``starting events'',
for which the earliest observed pulses occur within the instrumented volume.
The use of the outermost DOMs as a veto layer allows the rejection of
atmospheric muons that enter the detector from above or merely pass
sufficiently close to the instrumented volume to produce a signal capable of
surviving the initial filter.  Analyses of starting events are able to
accept neutrinos of all flavors and interaction types because only neutrinos
can pass undetected through the veto layer before interacting in the ice.
Veto methods currently used in IceCube analyses significantly reduce the
effective volume for detecting $\nu_\mu$-induced muon tracks, resulting in a
smaller final sample that is dominated by cascades.  The angular uncertainty
of cascade reconstructions ($\gtrsim10^\circ$) is large compared to that of
track reconstructions ($\lesssim1^\circ$).  However, the requirement that
the neutrino interaction vertex is located within the instrumented volume
results in good energy resolution (within $\sim10\%$)
\citep{Aartsen:2013vja} compared to incoming muon tracks originating at an
unknown distance from the detector.

Lower energy muons not only deposit less energy overall, but they may travel
larger distances between substantial energy losses due to the stochastic
nature of radiative processes at these energies.  The initial discovery of
astrophysical neutrinos used only the outermost DOMs as a veto layer and
thus achieved adequate background rejection only at energies
$\gtrsim\unit[60]{TeV}$ \citep{Aartsen:2014gkd}.  In a follow-up analysis,
the energy threshold was reduced to $\sim\unit[1]{TeV}$ by scaling the veto
thickness as a function of total collected charge \citep{Aartsen:2014muf}
such that events depositing as little as \unit[100]{PE} could be observed,
provided that the earliest light was found in the DeepCore in-fill array.

This adaptive veto event selection was applied to two years of data taken
from May 2010 to May 2012 | one year using the nearly-complete 79-string
configuration and one year using the complete 86-string detector.  In a
total of 641 days of IceCube livetime, 283 cascade and 105 track events were
found \citep{Aartsen:2014muf}.  While most of the track events are accepted
by $\nu_\mu$ point source searches \citep{Aartsen:2014cva} and a small
fraction of the cascades are included in the earlier high energy starting
event analysis \citep{Aartsen:2013jdh,Aartsen:2014gkd}, the majority of
these cascades have not yet been studied in the context of spatial
clustering.  In this paper, we turn our attention to 263 of these cascades
with deposited energies of $\unit[1]{TeV}-\unit[1]{PeV}$ to perform an
astrophysical neutrino source search that is complementary to and
statistically independent from traditional track analyses.

\begin{figure*}[t]
  \begin{center}
    \includegraphics[width=\columnwidth]{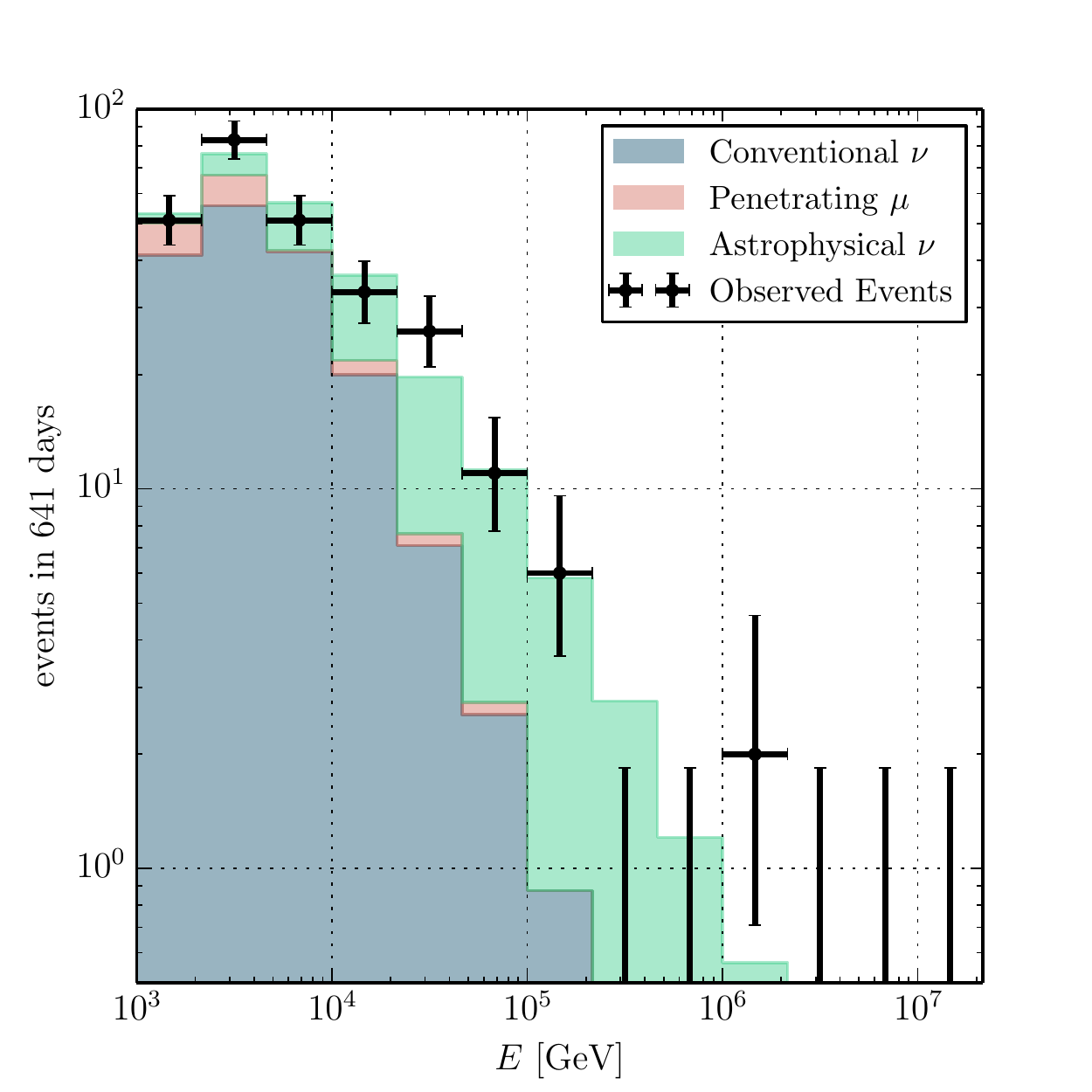}
    \hfill
    \includegraphics[width=\columnwidth]{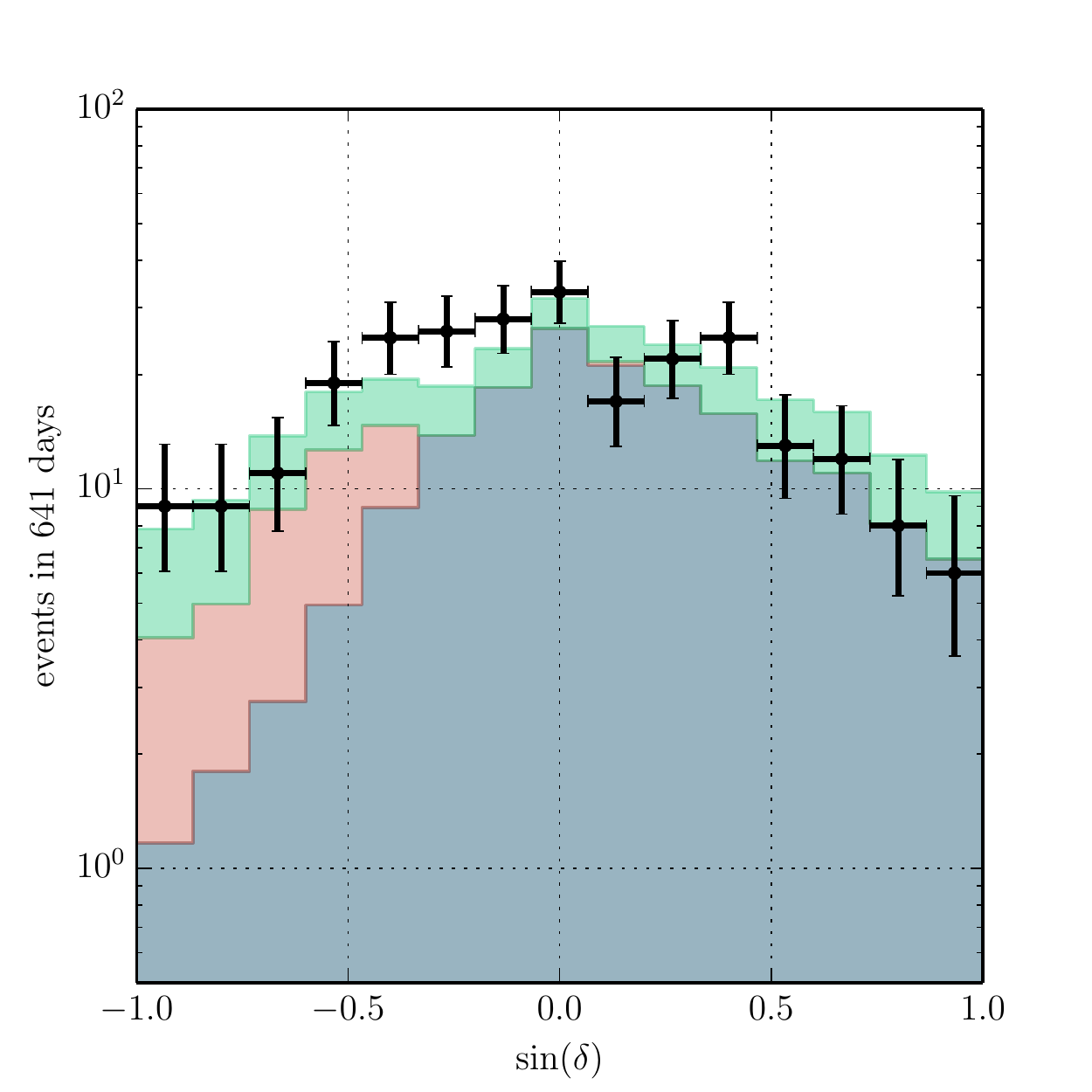}
  \end{center}
  \caption{Reconstructed energy (left) and declination (right) distributions
    for the best-fit atmospheric and astrophysical spectra (shaded
    histograms) obtained in \citet{Aartsen:2014muf} compared to the
    distributions for the 263 cascades (black crosses) depositing at least
    $\unit[1]{TeV}$ observed in that analysis.  Atmospheric muons
    misidentified as cascades after passing undetected through the veto
    layer are concentrated at $\sin(\delta)<-0.3$, while in the same range
    some atmospheric neutrinos are rejected because they are accompanied by
    incoming muons.
  }
  \label{fig:recodists}
\end{figure*}

\begin{figure}[t]
  \begin{center}
    \includegraphics[width=\columnwidth]{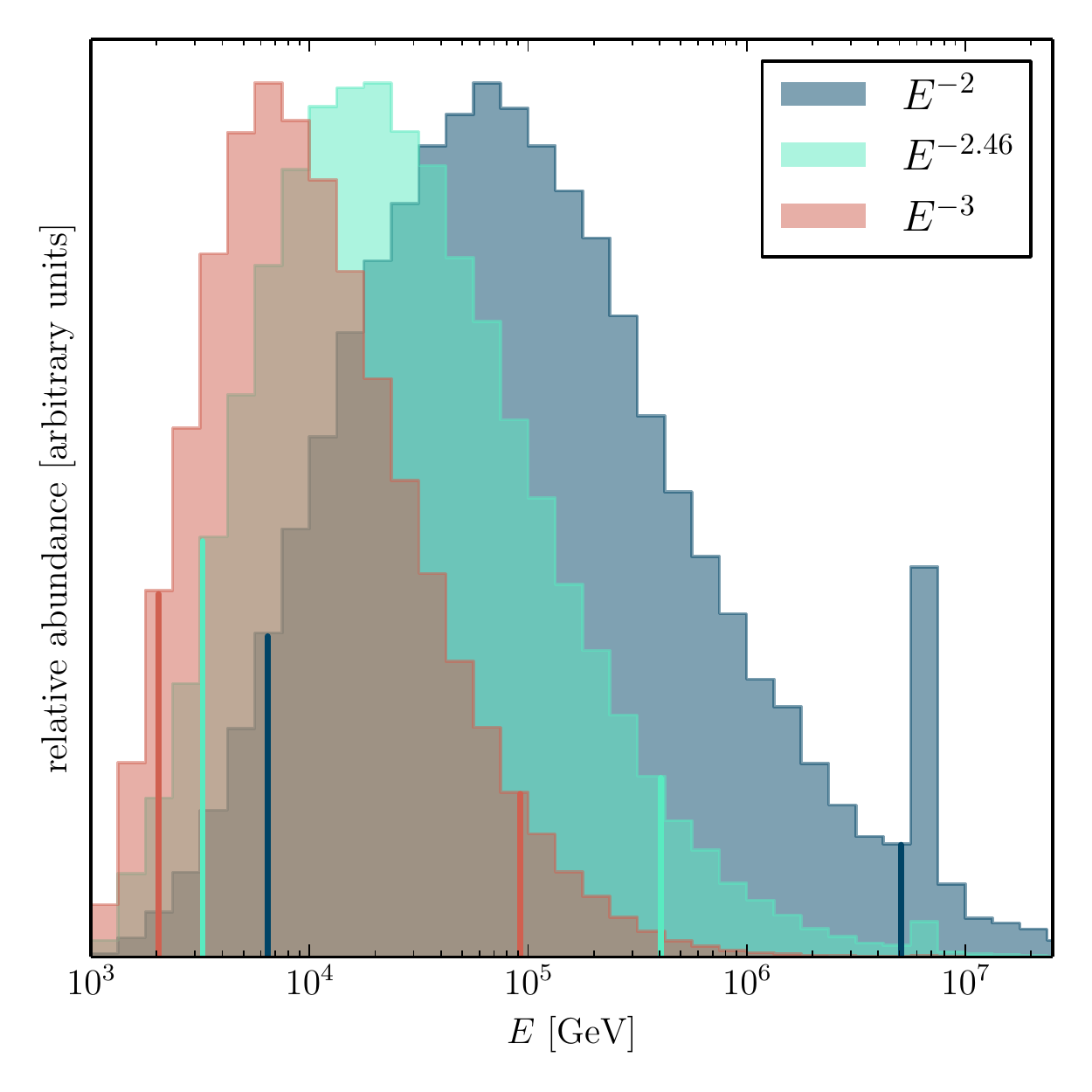}
  \end{center}
  \caption{Neutrino energy distributions expected from sources emitting a
    hard ($E^{-2}$, blue) or soft ($E^{-3}$, red) spectrum compared with the
    best-fit all-sky astrophysical component following an $E^{-2.46}$
    spectrum (green).  Each distribution includes all standard model
    neutrino flavors, assuming a 1:1:1 flavor ratio at Earth with equal
    fluxes of $\nu$ and~$\bar\nu$.  Vertical lines indicate intervals
    containing 90\% of events.  While no such events have yet been observed,
    an enhanced acceptance is expected for $\nu_e$ at $\unit[6.3]{PeV}$ due
    to the Glashow resonance~\citep{Glashow:PhysRev.118.316}.
  }
  \label{fig:Edist}
\end{figure}

The reconstructed energy and declination distributions for the 263 observed
cascades are compared in Figure~\ref{fig:recodists} with the expectation
from the best-fit atmospheric and astrophysical fluxes found in the spectral
analysis.  The fitted astrophysical component follows an $E^{-2.46}$
spectrum and contributes an expected $71^{+9.5}_{-8.4}$ cascades in 641 days
| a far larger fraction of the total event rate than in previous source
searches with tracks~\citep{Aartsen:2016oji}.  The neutrino energy
distribution is shown in Figure~\ref{fig:Edist} for the best-fit spectrum as
well as the hard ($E^{-2}$) and soft ($E^{-3}$) source spectrum hypotheses
tested directly in this paper.  For an $E^{-3}$ spectrum, we expect 90\% of
events to have energies between \unit[2]{TeV} and \unit[90]{TeV}; for an
$E^{-2}$ spectrum this range shifts to $\unit[6]{TeV}-\unit[5]{PeV}$.

\begin{figure}[t]
  \begin{center}
    \includegraphics[width=\columnwidth]{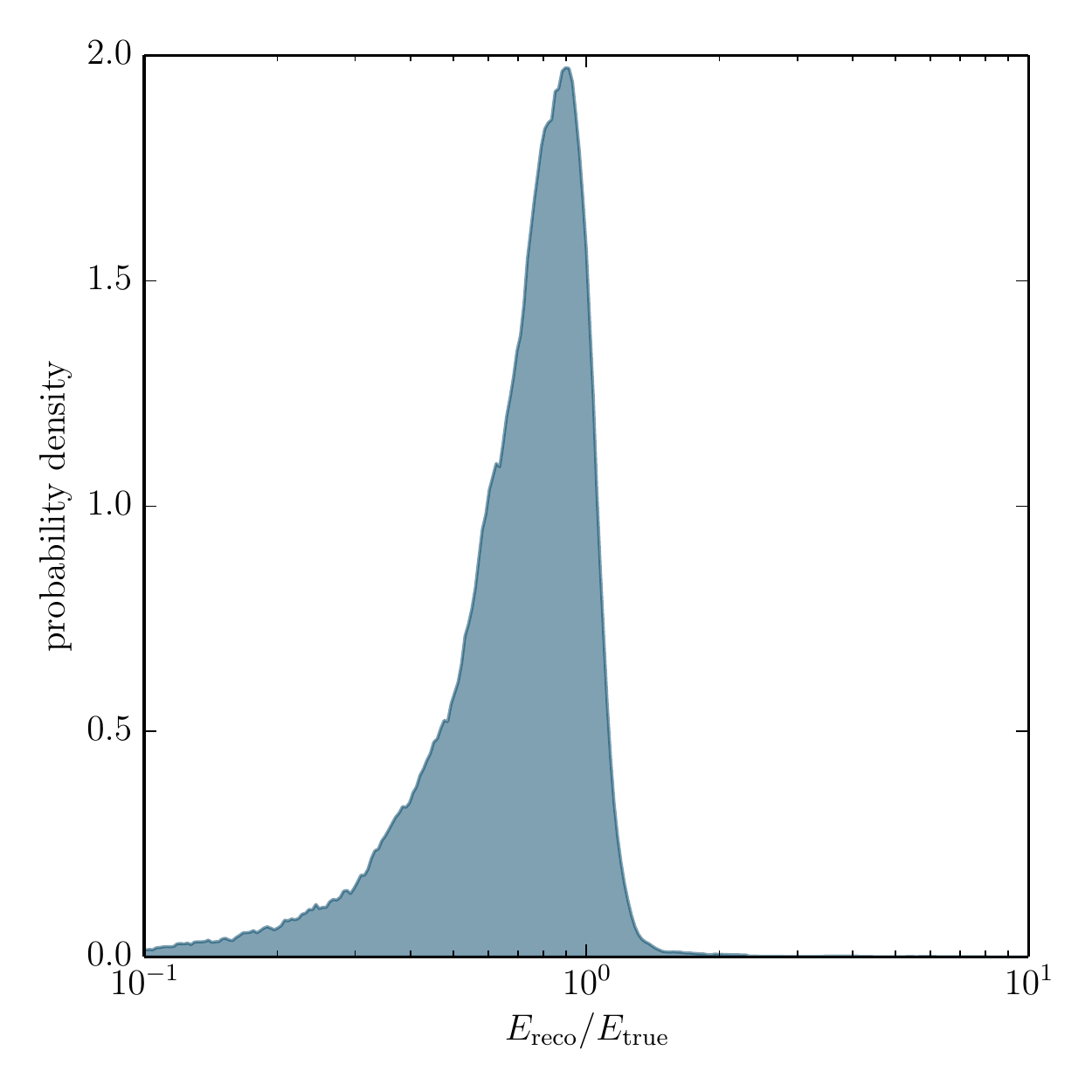}
  \end{center}
  \caption{Ratio of reconstructed to true neutrino energy for signal MC
    following an $E^{-2.46}$ spectrum.  Reconstructed energy is on average
    proportional to true neutrino energy for all interaction flavors, with
    agreement within $\sim10\%$ for 68\% of CC $\nu_e$ interactions.
  }
  \label{fig:res1D:energy}
\end{figure}

\begin{figure}[t]
  \begin{center}
    \includegraphics[width=\columnwidth]{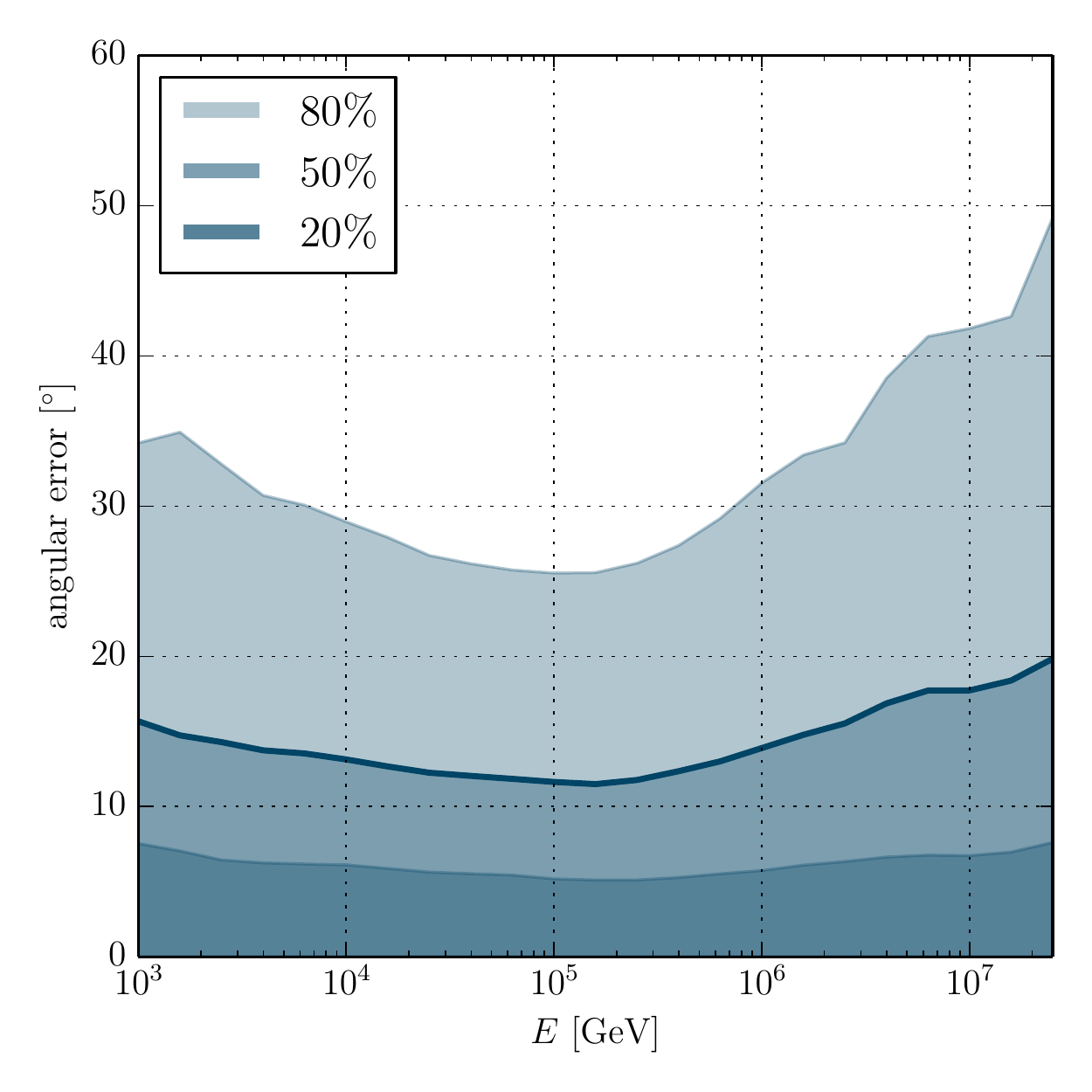}
  \end{center}
  \caption{Expected angular reconstruction performance as a function of
    neutrino energy.  Shaded regions indicate the radii of error circles
    covering 20\%, 50\%, and 80\% of events.  Below $\unit[20]{PeV}$, the
    median angular error, highlighted by the dark blue curve, ranges from
    $11^\circ$ to $20^\circ$.
  }
  \label{fig:angres}
\end{figure}

\begin{figure}[t]
  \begin{center}
    \includegraphics[width=\columnwidth]{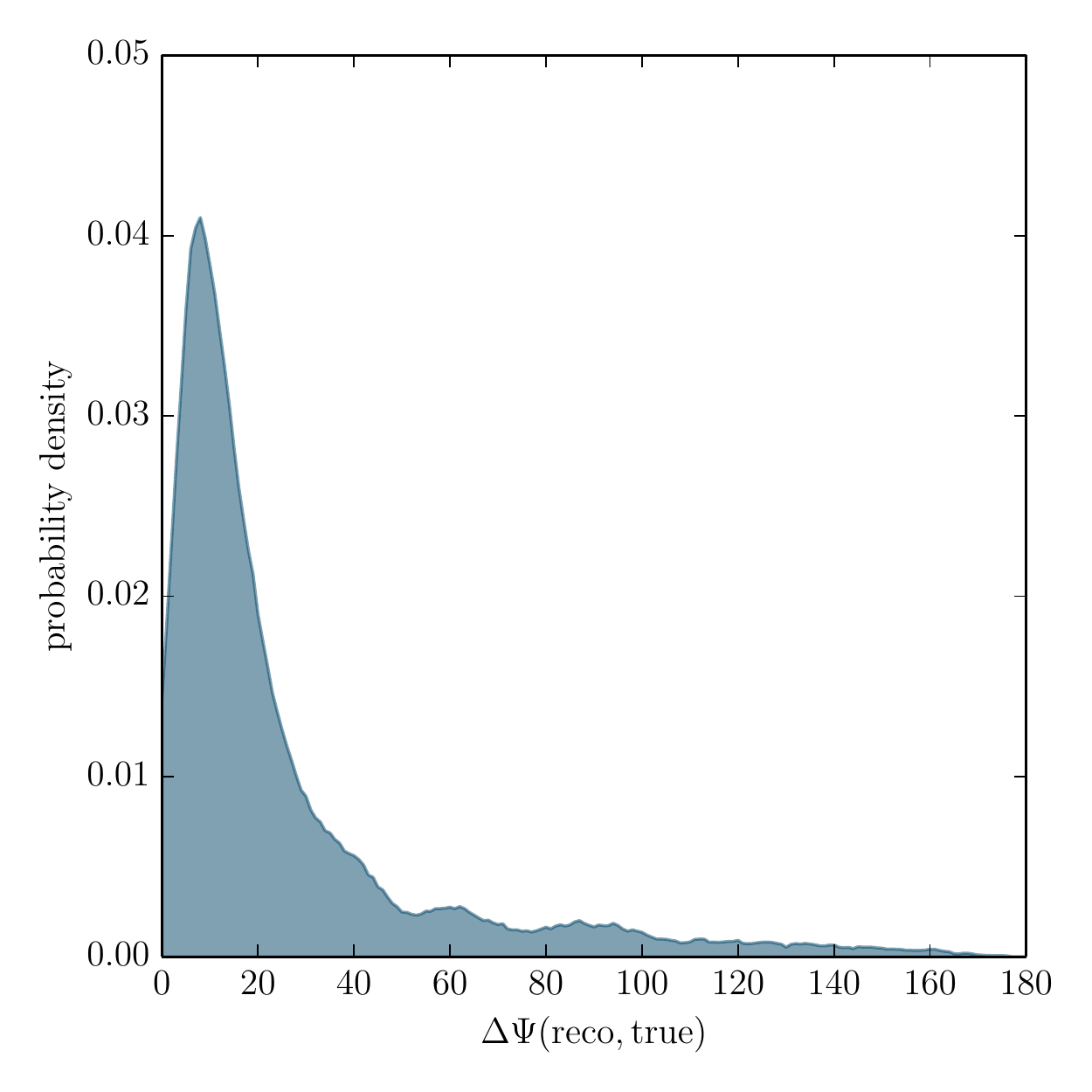}
  \end{center}
  \caption{Expected distribution of angular separation between reconstructed
    and true neutrino direction for signal MC following an $E^{-2.46}$
    spectrum.  While the distribution includes a tail extending all the way
    to $180^\circ$, 50\% (90\%) of events are reconstructed within
    $13^\circ$ ($45^\circ$).
  }
  \label{fig:res1D:angular}
\end{figure}

In this work, we use the same per-event reconstructions as in the spectral
analysis.  The strict containment requirement results in good energy
resolution up to at least $\sim\unit[\text{few}]{PeV}$.  The reconstructed
energy agrees with the neutrino energy within $\sim10\%$ for 68\% of CC
$\nu_e$ interactions and is on average proportional to neutrino energy for
other interaction flavors \citep{Aartsen:2014muf}.  Agreement between
reconstructed and true neutrino energy is shown in
Figure~\ref{fig:res1D:energy}.  The primary challenge for source searches
with cascades is the angular reconstruction, for which the performance is
shown as a function of energy in Figure~\ref{fig:angres} and averaged over
all energies in Figure~\ref{fig:res1D:angular}.  At low energies,
the reconstruction benefits to some degree from the preferential selection
of interactions in or near the more densely instrumented DeepCore.  At high
energies, performance is somewhat poorer than optimal | compare with, e.g.,
\citet{Aartsen:2013vja} | likely due to the specific reconstruction settings
used for this sample, which are less computationally intensive but which
employ a coarser description of the expected light yield and a less rigorous
scan of the directional likelihood landscape.

\newpage
\section{Methods and Performance}
\label{sec:methods}

We use an unbinned maximum likelihood method to quantify the extent to which
the observed events are more consistent with a spatially localized
astrophysical signal hypothesis than a randomly distributed background
hypothesis.  This method exploits the spatial distribution of events as well
as the distribution of per-event deposited energies, where the latter
improves the sensitivity to sources with harder spectra than atmospheric
backgrounds.  While we largely follow the approach used in traditional track
analyses \citep[most recently][]{Aartsen:2016oji}, the specific signal and
background models are modified to accommodate the large angular uncertainties
and overall low statistics of the cascade event selection.  In
Section~\ref{sec:likelihood} we review the likelihood construction,
including explanations for changes with respect to previous work with
tracks.  In Section~\ref{sec:hyps} we introduce the specific hypothesis
tests considered in this work.  Systematic uncertainties are discussed in
Section~\ref{sec:sys} and the performance of the cascade analysis is
presented in Section~\ref{sec:perf}.

\newpage
\subsection{Maximum Likelihood Method}
\label{sec:likelihood}

The likelihood is expressed as a product over events $i$:
\begin{align}
  \mathcal{L}(n_s, \gamma)
  &= \prod_i
  {} \Gb{\frac{n_s}{N}
  {} \mathcal{S}_i
  {} + \Gp{1 - \frac{n_s}{N}}
  {} \mathcal{B}_i},
  \label{eq:likelihood}
\end{align}
where $n_s$ is the number of signal events, $\gamma$ is the spectral index
of the source, $N=263$ is the total number of events, $\mathcal{S}_i$ is the
likelihood of event $i$ contributing to the source, and $\mathcal{B}_i$ is
the likelihood of event $i$ contributing to atmospheric or unresolved
astrophysical backgrounds.  $\mathcal{S}_i$ depends on the properties of
both event~$i$ and the source hypothesis (including spectral index
$\gamma$), while $\mathcal{B}_i$ depends only on the properties of the
events.  $\hat n_s$ and $\hat\gamma$ are the values that give the maximum
likelihood $\hat{\mathcal{L}}$, subject to the constraint that $\hat
n_s\geq0$.  Events that are more correlated spatially or energetically with
the source hypothesis obtain larger values for $\mathcal{S}_i$, driving the
fit towards larger values of $\hat n_s$ and $\hat{\mathcal{L}}$.

We approximate the signal and background likelihoods $\mathcal{S}_i$ and
$\mathcal{B}_i$ as products of space and energy factors: $S_i^\text{space}
\cdot S_i^\text{energy}$ and $B_i^\text{space} \cdot B_i^\text{energy}$,
respectively.  Each factor is obtained by convolving the properties of the
event origin | either astrophysical source, or atmospheric or unresolved
astrophysical background | first with the detector response and then with
the event reconstruction resolution.  For $B_i^\text{space}$, this is done
using a normalized histogram of reconstructed declination $\delta$ for an
ensemble of background-like events, accounting for detector effects and
smearing from finite angular resolution simultaneously.  Similarly, for
$S_i^\text{energy}$ and $B_i^\text{energy}$, we use normalized histograms of
the logarithm of the deposited energy $\log_{10}E$ for ensembles of
signal-like and background-like events, respectively, accounting for the
declination dependence with separately-normalized histograms in each of ten
bins in $\sin\delta$.  $S_i^\text{energy}$ is computed from signal Monte
Carlo (MC) on a grid of spectral indices $\gamma$ ranging from 1 to 4.  For
a given event, $B_i^\text{space}$, $S_i^\text{energy}$ and
$B_i^\text{energy}$ are equal to the values of the histograms for the bin
containing the event.  The location of IceCube at the geographic South Pole
allows us to express these factors as functions of declination, rather than
zenith angle with respect to the detector, without loss of information.  A
small additional dependence on azimuth angle is neglected.

In the classic track analysis, the background per-event likelihoods are
constructed from the full experimental dataset.  With a large sample of
well-reconstructed muon tracks dominated by atmospheric backgrounds, both
$B_i^\text{space}$ and $B_i^\text{energy}$ are well constrained
statistically even for dense binning in both $\sin\delta$ and $\log_{10}E$.
By contrast, our sample of only 263 cascade events is only sufficient to
constrain $B_i^\text{space}$.  Thus our first modification to the method is
to construct $B_i^\text{energy}$ from neutrino and atmospheric muon MC
simulations weighted to the best-fit atmospheric and astrophysical spectra
found by the all-sky flux analysis using these events
\citep{Aartsen:2014muf}.  In this way we obtain a detailed estimate of the
energy distribution throughout the sky, even at energies not yet observed at
all declinations in two years of experimental livetime.

The signal space factor $S_i^\text{space}$ is obtained by convolving a
source hypothesis with an analytical estimate of the spatial probability
density distribution for event $i$ originating at reconstructed right
ascension and declination $(\alpha_i,\delta_i)$.  In track analyses, it is a
good approximation to model this distribution as a 2D Gaussian with width
$\sigma_i$ estimated event-by-event using a dedicated reconstruction.  We
modify this treatment for cascades both because the angular uncertainties
are much larger and because it is too computationally expensive to estimate
them directly for each event.

In this analysis we parameterize the angular resolution as a function of
reconstructed declination $\delta_i$ and energy $E_i$.  In parts of this
parameter space, either the declination or right ascension errors tend to be
systematically larger, so these are treated independently.  For each of 10
bins in $\sin\delta$ and 12 in $\log_\text{10}E$, we find the values
$\sigma_\alpha$ and $\sigma_\delta$ such that
$|\alpha_i-\alpha_i^\text{true}|<\sigma_\alpha$, and separately
$|\delta_i-\delta_i^\text{true}|<\sigma_\delta$, for 68.27\% of simulated
events in the bin.  The spatial probability density distribution for
observed event $i$ is the product of 1D Gaussians with these widths,
normalized such that the distribution integrates to unity on the sphere.

We consider two types of source hypothesis: point sources and the galactic
plane | an extremely extended source.  A point source is modeled as a 2D
delta distribution centered at the source coordinates.  The expected
emission from the galactic plane is in general model-dependent.  Here we
represent the galactic plane as a simple line source at galactic latitude
$b=0$.  In either case, $S_i^\text{space}$ is obtained by convolving the
source hypothesis with the per-event spatial probability density
distributions described above.  For point sources, the convolution is
trivial; for the galactic plane, it is evaluated numerically on a grid with
$1^\circ$ spacing.

\subsection{Hypothesis Tests}
\label{sec:hyps}

In this work we consider three search categories: (1)~a scan for point-like
sources anywhere in the sky, (2)~a search for neutrinos correlated with an
\emph{a priori} catalog of promising source candidates, and (3)~a search for
neutrinos correlated with the galactic plane.  Each search entails multiple
specific hypothesis tests.  The all-sky scan tests for point-like sources on
a dense grid of coordinates throughout the sky.  The catalog search tests
the coordinates of each source candidate individually.  The galactic plane
search includes partially correlated tests for a hypothesis including the
entire galactic plane and a hypothesis including only the part of the
galactic plane in southern sky.

\newcommand{\calT}{\mathcal{T}}
The test statistic used to compute significances is the likelihood ratio:
\begin{align}
  \calT
  &= -2\ln\Gb{\frac{\mathcal{L}(n_s=0)}{\mathcal{L}(\hat n_s, \hat\gamma)}},
\end{align}
where $\mathcal{L}(n_s=0)$ is the background-only likelihood and is
independent of $\gamma$.  For an individual hypothesis, the pre-trials
significance $p_\text{pre}$ of an observation yielding a test statistic
$\calT_\text{obs}$ is the probability of observing $\calT>\calT_\text{obs}$
if the background-only hypothesis were true.  The background-only $\calT$
distribution is found by performing the likelihood test on a large number of
ensembles with randomized $\alpha_i$, which removes any clustering that may
be present in the true event ensemble.  At declinations close to the poles,
$|\delta_i|>60^\circ$, randomizing $\alpha_i$ alone is insufficient to
remove a possible cluster of cascades.  This is addressed by additionally
randomizing $\sin\delta_i$ for the 15 events within these regions.

The pre-trials results, $p_\text{pre}$, do not account for multiple and
partially correlated hypothesis tests conducted in each search category.
The post-trials significance is determined by the most significant
$p_\text{pre}$ for any hypothesis in the category.  Specifically, for each
search category we find the post-trials probability $p_\text{post}$ of
observing any $\min(p_\text{pre})<\min(p_\text{pre})_\text{obs}$ if the
background-only hypothesis were true.  The background-only
$\min(p_\text{pre})$ distribution is found by generating additional
randomized event ensembles and noting the most significant $p_\text{pre}$ in
each one.  This construction leads to one final significance $p_\text{post}$
for each type of search; a further look-elsewhere effect between the
all-sky, source candidate catalog, and galactic plane searches is not
explicitly accounted for.  This method is conservative in that it strictly
controls only the false positive, but not the true positive, error rate.

We use the classical statistical approach \citep{neyman,lehmann2005testing}
to calculate the sensitivity, discovery potential, and flux upper limits.
The flux level is determined using randomized trials in which signal MC
events are injected at a Poisson rate $n_\text{sig}$ and distributed
according to the spatial and energetic properties of the signal hypothesis.
The remaining $N-n_\text{sig}$ events are injected according to the
background modeling procedure described above.  The sensitivity flux is that
which gives a 90\% probability of obtaining $\calT>\calT_\text{med}$, where
$\calT_\text{med}$ is the median of the background-only $\calT$
distribution.  The discovery potential flux is obtained by the same
procedure, but for a 50\% probability of yielding a $5\sigma$ pre-trials
significance.  The 90\% confidence level upper limit is the larger of either
the sensitivity or that flux which gives a 90\% probability of obtaining
$\calT>\calT_\text{obs}$.

\newpage
\subsection{Systematic Uncertainties}
\label{sec:sys}

The randomization procedures described in the previous section yield
background models and significances that are robust against systematic
uncertainties.  However, flux calculations in this analysis are based on
detailed neutrino signal MC as described in \citet{Aartsen:2016xlq} and are
subject to systematic uncertainties.  We estimate the impact of these
uncertainties on our results via their impact on the cascade angular
resolution and signal acceptance.  Of these, uncertainties related to the
angular resolution are the dominant effect.  Reconstruction performance
estimates from the baseline MC are limited by statistical uncertainties in
the observed light as well as any practical computational tradeoffs made in
data processing.  These estimates do not account for possible systematic
errors in the modeling of light absorption and scattering in either the bulk
of South Pole glacial ice or the narrow columns of refrozen ice surrounding
the DOMs.  Uncertainties in the light yield from showers and the optical
efficiency of the DOMs are also neglected in the baseline MC.  Taken
together, we estimate that these effects introduce an angular resolution
uncertainty that can be approximated as a Gaussian smearing of the baseline
point spread function with width $\sigma_\text{sys}\sim8^\circ$ (compare,
  e.g., the typical per-event errors in \citet{Aartsen:2014gkd} with the
median expected pure-statistical errors in \citet{Aartsen:2013vja}).
Applying this smearing weakens the sensitivity by $\sim20\%$ ($\sim23\%$)
for sources following an $E^{-2}$ ($E^{-3}$) spectrum, approximately
independent of source declination.

The uncertainties described above also have a small impact on the estimated
signal acceptance of the event selection.  Uncertainties in the DOM
efficiency are on average inversely correlated with uncertainties in the
scattering and absorption coefficients, so we can safely estimate the impact
of these uncertainties using a parameterization from available MC datasets
which only vary the DOM efficiency explicitly.  We consider a reduced DOM
efficiency of $-10\%$ relative to the baseline MC, which decreases both the
number of accepted events for a given flux and the reconstructed deposited
energy of each simulated event.  Under this change most signal events are
assigned slightly smaller weights $(S/B)_i^\text{energy}$ and some fall
below the detection threshold, weakening the sensitivity by $\sim4\%$,
approximately independent of source spectrum and declination.

The signal acceptance also depends on the neutrino interaction cross
section, which is known within a similar uncertainty $+4\%/-2.4\%$ below
\unit[100]{PeV} \citep{CooperSarkar:2011pa}.  The resulting impact on this
analysis is in general dependent on declination and neutrino energy, as an
increased (decreased) cross section would simultaneously increase (decrease)
the probability of detecting a neutrino upon arrival in the instrumented
volume but decrease (increase) the probability of a neutrino reaching the
detector after passing through the intervening earth and ice.  We take
$\sim4\%$ as a conservative estimate of the acceptance uncertainty due to
neutrino interaction cross section uncertainties.

While the signal acceptance depends largely on the total amount of light
recorded by the DOMs, the angular resolution depends most strongly on the
spatial and temporal distribution of light in the detector.  Therefore, we
take these effects to be approximately independent and add the above values
in quadrature to obtain a total systematic uncertainty of 21\% (24\%) for
sources following an $E^{-2}$ ($E^{-3}$) spectrum.  All following
sensitivities, discovery potentials, and flux upper limits include this
factor.

%\newpage
\subsection{Performance}
\label{sec:perf}

\begin{figure}[t]
  \begin{center}
    \includegraphics[width=\columnwidth]{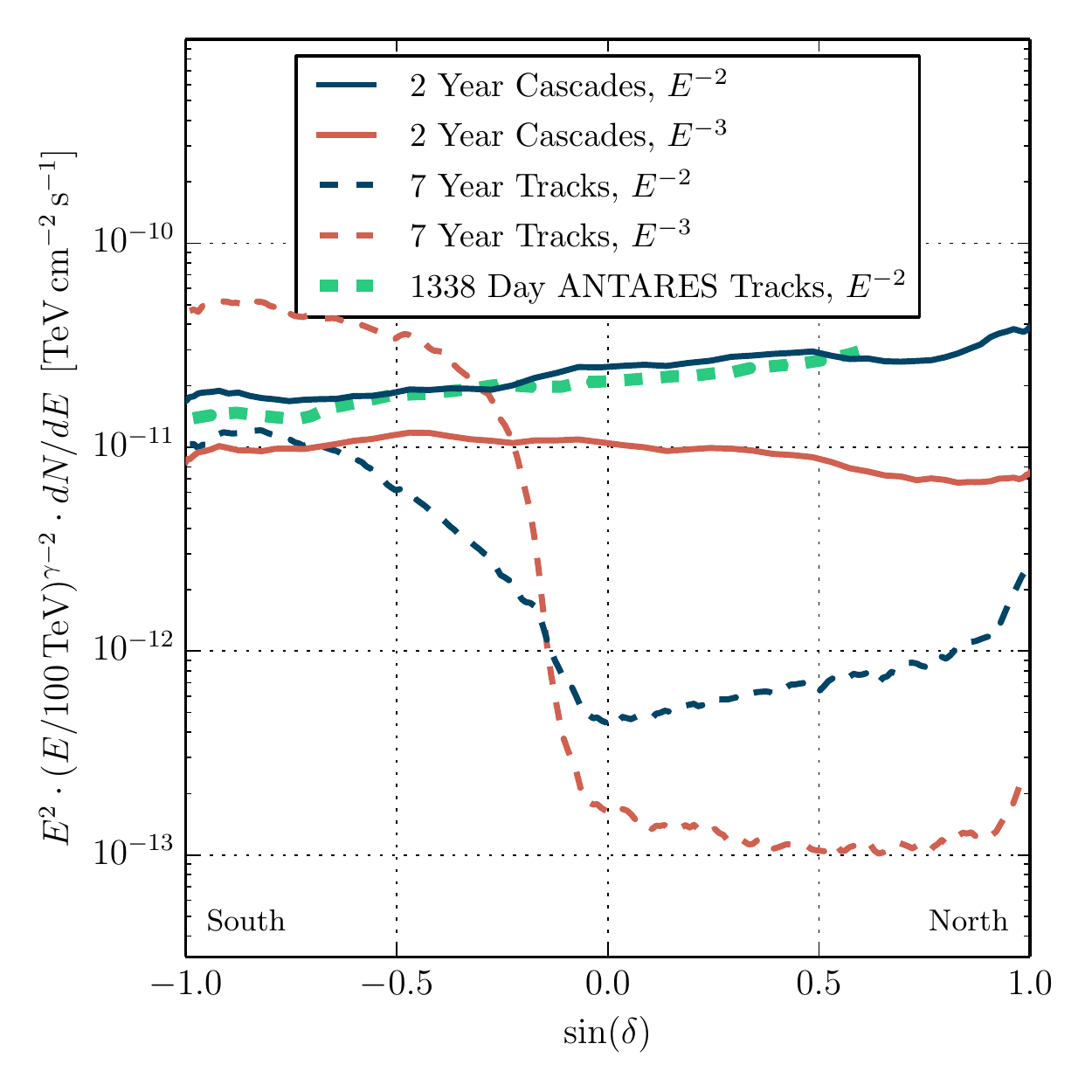}
  \end{center}
  \caption{Per-flavor sensitivity of the present 2-year cascade analysis and
    previous 7-year IceCube \citep{Aartsen:2016oji} and 1338-day ANTARES
    \citep{Adrian-Martinez:2014wzf} track analyses as a function of
    declination for a hard spectrum ($\gamma=2$) and soft spectrum
    ($\gamma=3$).
  }
  \label{fig:sens}
\end{figure}

\begin{figure}[t]
  \begin{center}
    \includegraphics[width=\columnwidth]{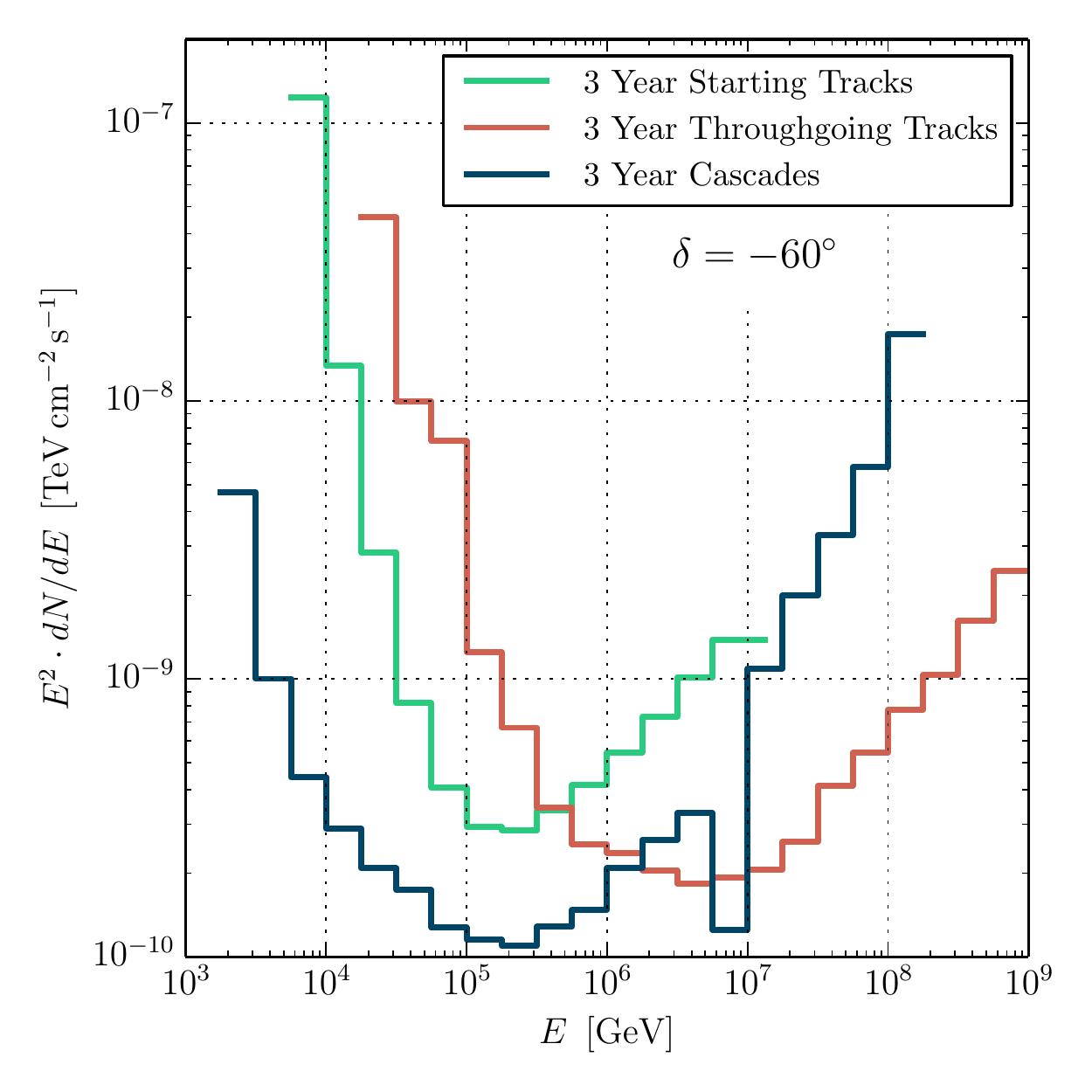}
  \end{center}
  \caption{Per-flavor differential sensitivity for a source at
    $\delta=-60^\circ$ for track analyses of throughgoing
    \citep{Aartsen:2014cva} and starting \citep{Aartsen:2016tpb} tracks,
    compared to this cascade analysis using the event selection from
    \citet{Aartsen:2014muf}.  The sensitivity in cascades is enhanced at
    $\unit[6.3]{PeV}$ due to the Glashow resonance
    \citep{Glashow:PhysRev.118.316}.  In this plot, all sensitivities are
    calculated for an equal three year exposure.
  }
  \label{fig:diffsens}
\end{figure}

\begin{figure}[h!]
  \begin{center}
    \includegraphics[width=\columnwidth]{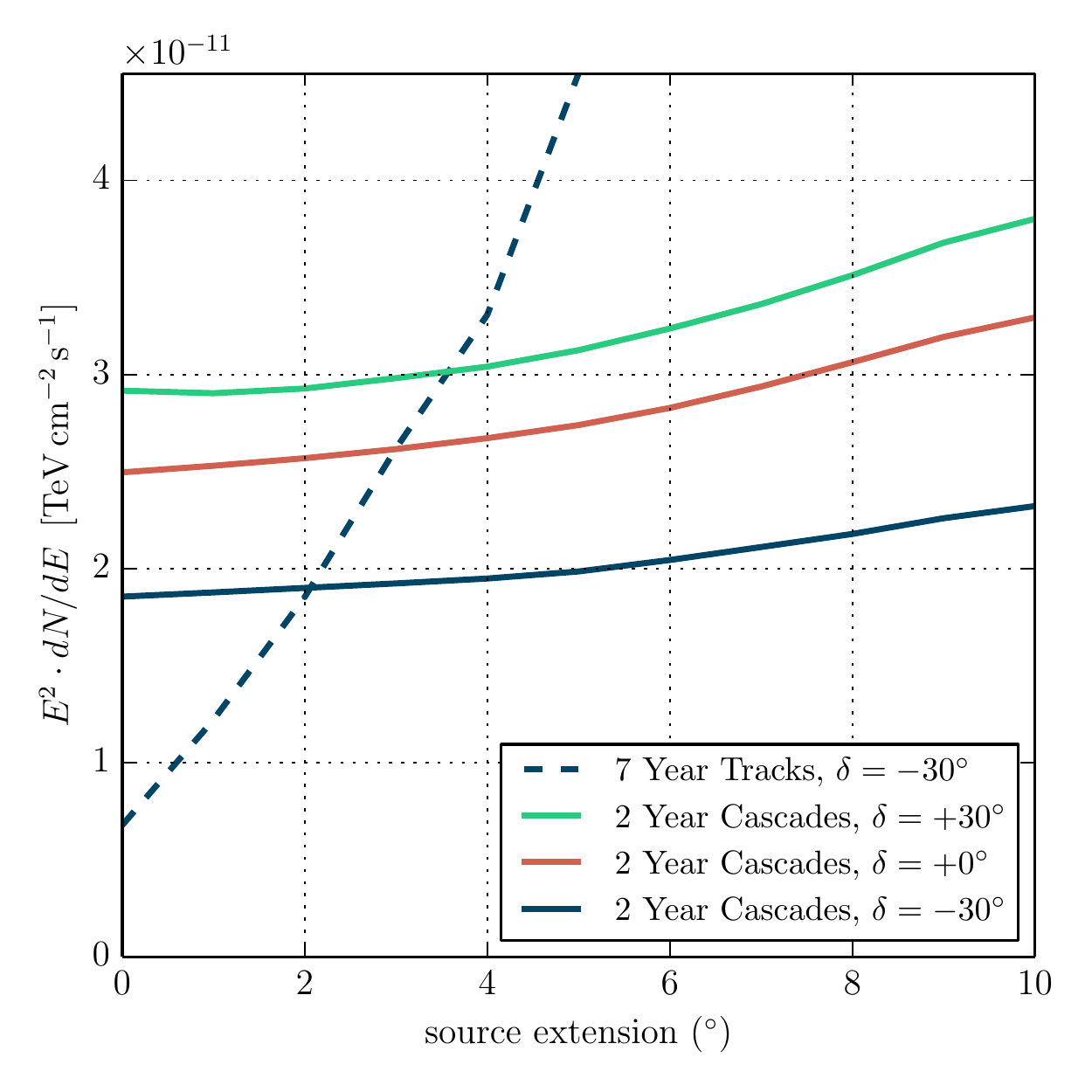}
  \end{center}
  \caption{Per-flavor sensitivity as a function of angular extension of the
    source.  For cascades, a point source hypothesis is used in the
    likelihood regardless of injected source extension.  For tracks, the
    sensitivity is found for an extended source hypothesis matching the
    injected signal using the throughgoing track dataset from
    \citet{Aartsen:2016oji}.
  }
  \label{fig:sensvext}
\end{figure}

The per-flavor sensitivity flux as a function of source declination for this
work and the most recently published IceCube \citep{Aartsen:2016oji} and
ANTARES \citep{Adrian-Martinez:2014wzf} track analyses are compared in
Figure~\ref{fig:sens}.  The cascade sensitivity shows only weak declination
dependence and, for an $E^{-2}$ spectrum, roughly traces the sensitivity of
ANTARES.  Near the South Pole, the sensitivity is enhanced by the veto of
atmospheric neutrinos accompanied by muons from the same cosmic ray-induced
shower.  The sensitivity is weaker near the horizon, where this veto of
atmospheric neutrinos is not possible.  From the horizon to the North Pole,
the sensitivity then improves for a soft $E^{-3}$ spectrum but continues to
weaken for a hard $E^{-2}$ spectrum because high-energy neutrinos are
subject to significant absorption in transit through the Earth.  The
sensitivity of the classic track search, by contrast, is strongly
declination-dependent, with best performance in the northern sky.  For a
southern source with a soft spectrum, the sensitivity flux is better with
just two years of cascades than with seven years of tracks.

We further explore the sensitivity to a southern source at
$\delta=-60^\circ$ in Figure~\ref{fig:diffsens}, which shows the per-flavor
sensitivity flux for an $E^{-2}$ signal spectrum injected in quarter-decade
bins in neutrino energy.  Here we directly compare the cascade and track
channels by scaling each analysis to an equal three year livetime | the same
exposure as in the first IceCube point source search to make use of starting
tracks \citep{Aartsen:2016tpb}.  At this declination, the low background
cascade search is more sensitive to such a southern source than IceCube
track-based searches up to $\sim\unit[1]{PeV}$.

Because of the large angular uncertainty for cascade events in IceCube, the
sensitivity depends only weakly on the angular size of the source.  In
Figure~\ref{fig:sensvext}, the sensitivity is shown as a function of angular
extension of the source.  The source extension is modeled as a Gaussian
smearing of a point source hypothesis.  For a smearing of up to $10^\circ$,
the sensitivity of this search is only 30\% weaker than for a point source.
In the classic track searches with angular resolution $\lesssim1^\circ$, the
sensitivity flux increases much more rapidly with source extension | even
when a matching extended source hypothesis is used in the likelihood.  As
shown in Figure~\ref{fig:sensvext}, the per-flavor sensitivity flux for a
source with extension $\geq2^\circ$ in the southern sky at
$\delta\leq-30^\circ$ is lower with just two years of cascades than with
seven years of tracks.  The cascade analysis performance is sufficiently
independent of source extension that we need not apply dedicated extended
source hypothesis tests in this work.

%\newpage
\section{Results}
\label{sec:results}

\begin{figure}[t]
  \begin{center}
    \vspace{1.3em}
    \includegraphics[width=\columnwidth]{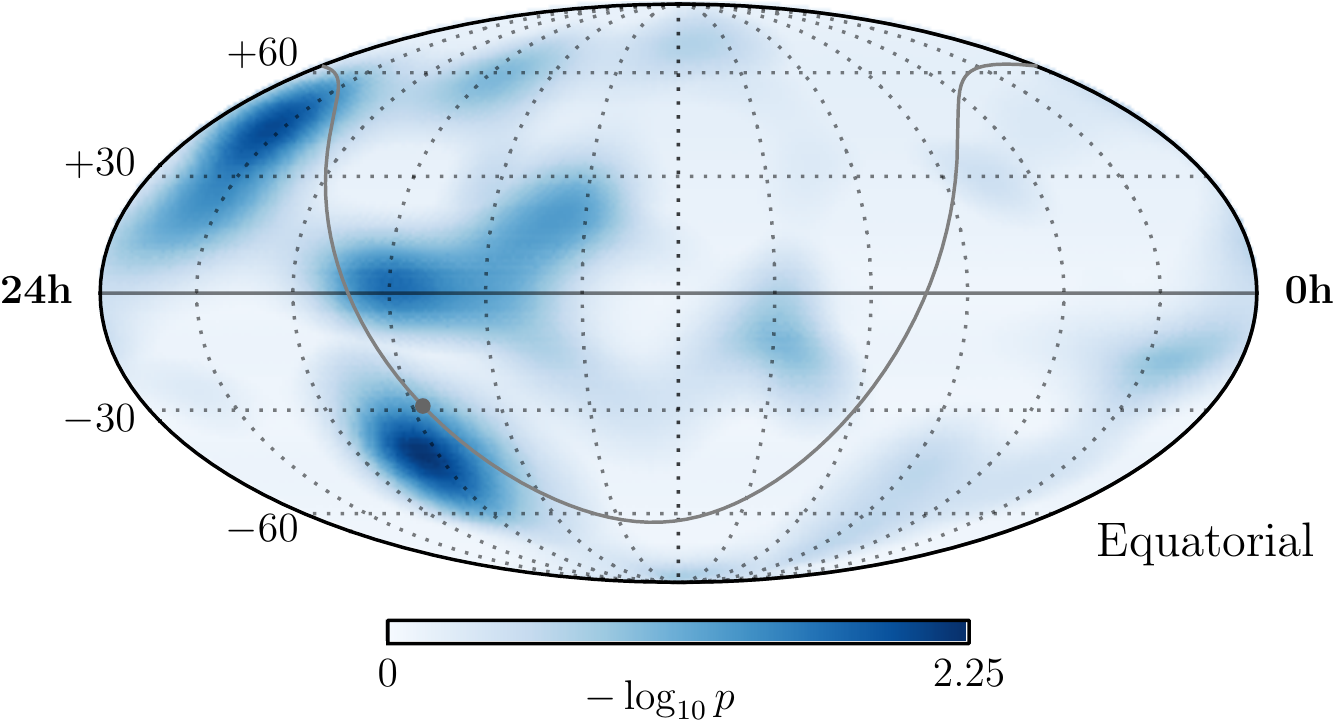}
  \end{center}
  \caption{Two-year starting cascade skymap in equatorial coordinates
  (J2000).  The skymap shows pre-trial $p$-values for all locations in the
  sky.  The grey curve indicates the galactic plane, and the grey dot
  indicates the galactic center.}
  \label{fig:skymap}
\end{figure}

The result of the all-sky scan is shown in Figure~\ref{fig:skymap}.  The
most significant deviation from the isotropic expectation is found in the
southern sky at $(\alpha,\delta)=(277.3^\circ,-43.4^\circ)$.  The pre-trials
significance is $p_\text{pre}=0.6\%$, and the best-fit number of signal
events and spectral index are $\hat n_s=7.1$ and $\hat\gamma=2.2$,
respectively.  Accounting for the large number of partially correlated
hypothesis tests in this scan, as described in \ref{sec:hyps}, the
post-trials significance is $p_\text{post}=66\%$.

\begin{figure*}[t]
  \begin{center}
    \includegraphics[width=\columnwidth]{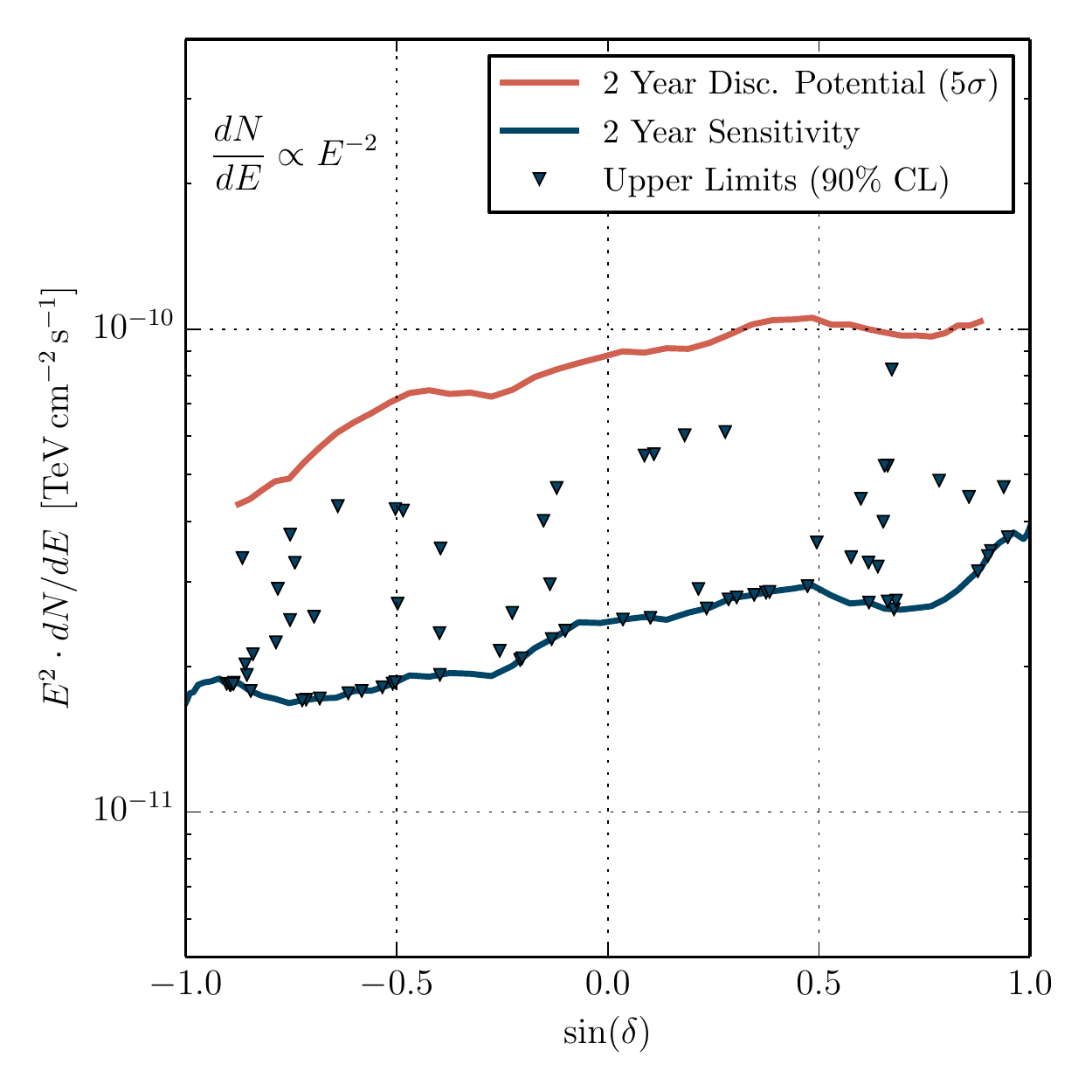}
    \hfill
    \includegraphics[width=\columnwidth]{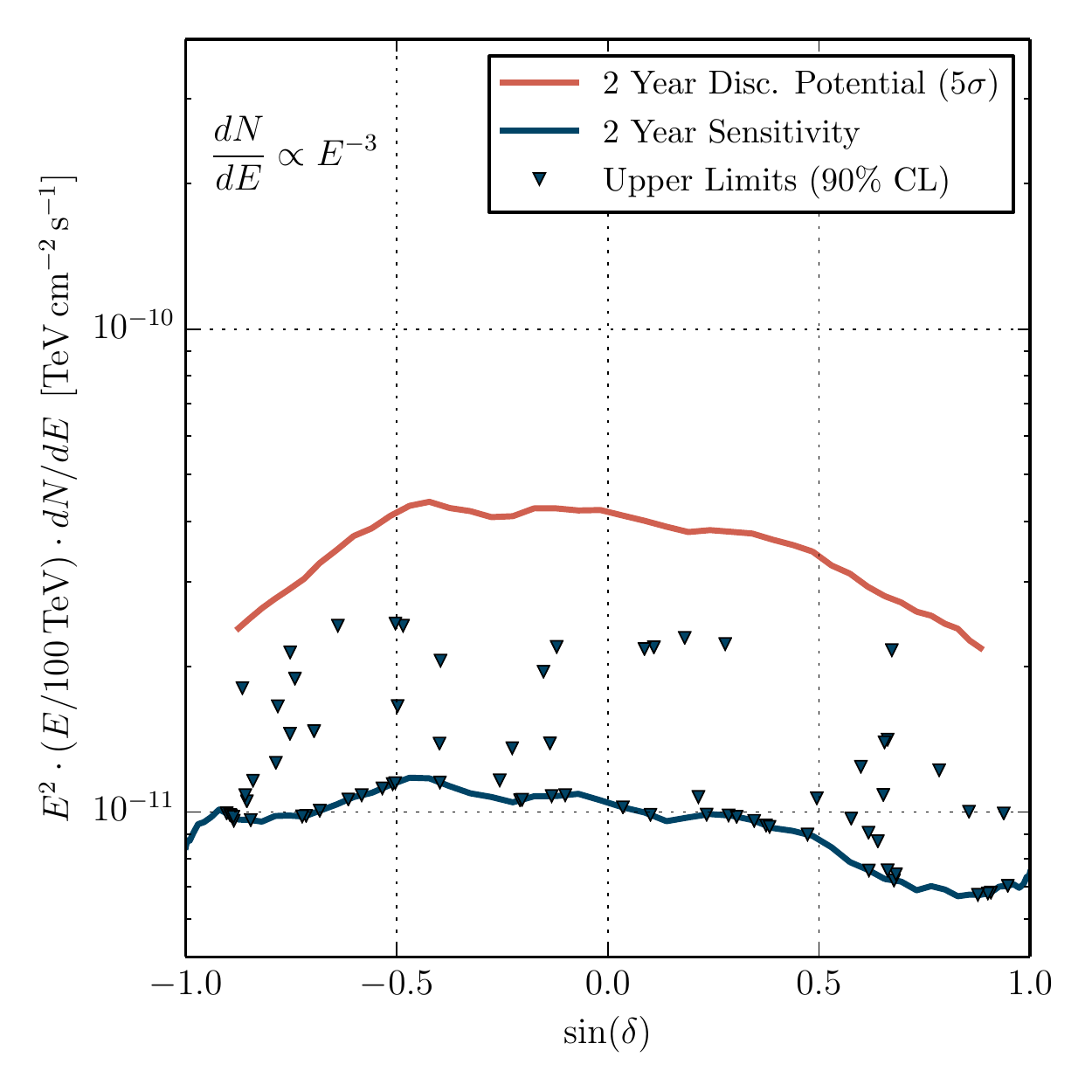}
  \end{center}
  \caption{Sensitivity and $5\sigma$ discovery potential as functions of
    declination, with flux upper limits for each object in the source
    catalog.  Left: hard spectral assumption ($\gamma=2$).  Right: soft
    spectral assumption ($\gamma=3$).
  }
  \label{fig:limits:E2}
\end{figure*}

For the source candidate catalog search, an ensemble of 74 promising source
candidates was selected \emph{a priori} by merging previously studied
catalogs of interesting galactic and extra-galactic objects
\citep{Aartsen:2016oji,Adrian-Martinez:2015ver}.  The result of the search
is shown in Table~\ref{tab:cat}.  The most significant source is BL Lac,
located at $(\alpha,\delta)=(330.68^\circ,42.28^\circ)$.  The pre-trials
significance is $p_\text{pre}=1.0\%$, and the best-fit number of signal
events and spectral index are $\hat n_s=6.9$ and $\hat\gamma=3.0$,
respectively.  The post-trials significance is $p_\text{post}=36\%$.  Flux
upper limits for each object in the catalog are shown in
Figure~\ref{fig:limits:E2} along with the sensitivity and $5\sigma$
discovery potential as functions of declination.

Of the galactic plane searches, the southern-sky-only hypothesis test was
more significant, with a pre-trials $p_\text{pre}=50\%$.  The fit obtained
$n_s=2.7$ and $\gamma=2$.  This test is strongly correlated with the all-sky
search; the post-trials significance is $p_\text{post}=65\%$.

%\newpage
\section{Conclusion and Outlook}
\label{sec:conclusion}

In this first search for sources of astrophysical neutrinos using cascades
with energies as low as $\unit[1]{TeV}$ in two years of IceCube data, no
significant source was found.  This result is consistent with previous
$\nu_\mu$ searches
\citep{Aartsen:2016oji,AdrianMartinez:2012rp,Adrian-Martinez:2015ver} which
already find stringent constraints on emission from astrophysical point
sources of neutrinos.  Nevertheless, this analysis shows that despite large
angular uncertainties, all-flavor source searches with cascades are
surprisingly sensitive, particularly to emission from southern sources that
follow a soft energy spectrum or are spatially extended.  This type of
analysis is therefore complementary to standard $\nu_\mu$ searches, which
are most sensitive to point-like and northern sources.

\begin{figure}[t]
  \begin{center}
    \includegraphics[width=\columnwidth]{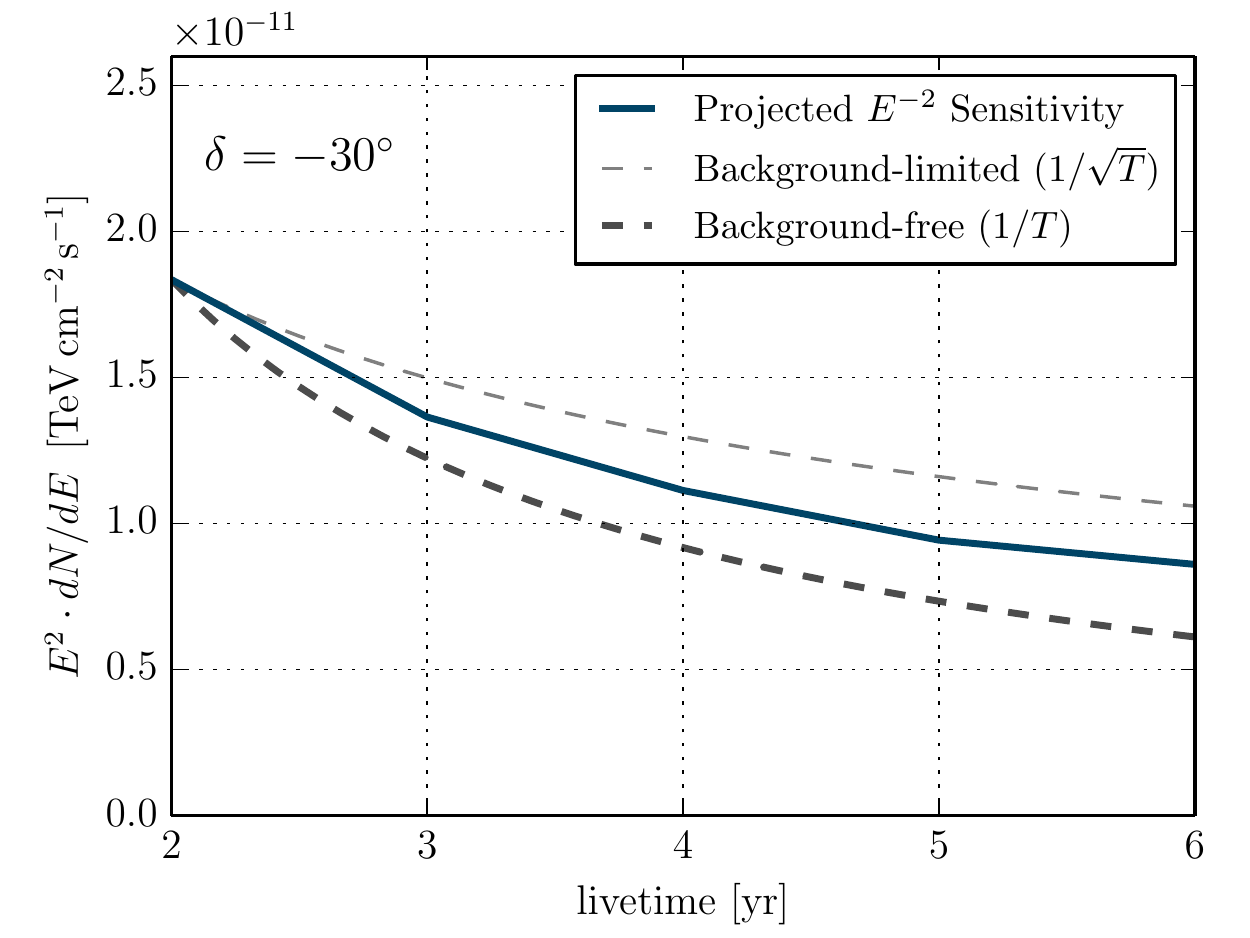}
  \end{center}
  \caption{Projected sensitivity as a function of detector livetime for a
  source at $\delta=-30^\circ$.  Time evolution scaling with $1/T$
  (background-free case) and $1/\sqrt{T}$ (background-limited case) are
  shown in thick and thin grey dashed curves, respectively.}
  \label{fig:sensvtime}
\end{figure}

Future source searches with cascades will benefit from several improvements.
Most importantly, the adaptive veto method will soon be applied to at least
four more years of IceCube data.  Because of the low background in this
event selection, the sensitivity strengthens faster than $[\text{detector
livetime}]^{-1/2}$, as shown in Figure~\ref{fig:sensvtime}.  Ongoing work on
the optimization of cascade angular reconstructions, including increasingly
detailed studies of Cherenkov light propagation in South Pole glacial ice,
may lead to angular resolution improvements that increase the cascade
channel signal-to-background ratio further still.

In this work, we searched for neutrino emission from a catalog of source
candidates previously studied in track analyses
\citep{Aartsen:2016oji,Adrian-Martinez:2015ver}.  The catalog was optimized
in light of the strengths of those analyses, and thus includes many northern
sources which would almost certainly be visible first in throughgoing
tracks.  We may be able to improve the discovery potential for future
catalog analyses with cascades by considering a catalog of source candidates
for which this analysis is best-suited, such as extended objects in the
southern sky.

We have considered only very simple models for extended emission from the
galactic plane, which we have treated here as a uniform line source.
However, detailed models \citep{Ackermann:2012:GammaProfile,Gaggero:2015xza}
have been constructed to account for the measured distribution of $\gamma$
emission from poorly resolved sources and cosmic ray interactions with
galactic dust clouds.  Future cascade analyses will test these models
directly, leading to clearer statements on neutrino emission within our
own galaxy.

Here we have searched only for steady, time-independent neutrino emission,
but the conclusions of this paper apply equally well to transient sources.
While a cascade event selection has been added to IceCube's gamma-ray burst
analysis \citep{Aartsen:2016qcr}, other time-dependent analyses
\citep[e.g.][]{Aartsen:2015wto} have not yet made use of this channel.  In
the future, searches for emission from objects such as flaring AGN could
benefit from the inclusion of neutrino-induced cascades.  Proposed
next-generation detectors \citep{Aartsen:2014njl,Adrian-Martinez:2016fdl}
may also benefit by considering source searches with the cascade channel in
the optimization of their optical sensors and array geometry.

%\input{acknowledgments}
% acknowledgments.tex

%\newpage
\acknowledgments

We acknowledge the support from the following agencies:
U.S. National Science Foundation-Office of Polar Programs,
U.S. National Science Foundation-Physics Division,
University of Wisconsin Alumni Research Foundation,
the Grid Laboratory Of Wisconsin (GLOW) grid infrastructure at the University of Wisconsin - Madison, the Open Science Grid (OSG) grid infrastructure;
U.S. Department of Energy, and National Energy Research Scientific Computing Center,
the Louisiana Optical Network Initiative (LONI) grid computing resources;
Natural Sciences and Engineering Research Council of Canada,
WestGrid and Compute/Calcul Canada;
Swedish Research Council,
Swedish Polar Research Secretariat,
Swedish National Infrastructure for Computing (SNIC),
and Knut and Alice Wallenberg Foundation, Sweden;
German Ministry for Education and Research (BMBF),
Deutsche Forschungsgemeinschaft (DFG),
Helmholtz Alliance for Astroparticle Physics (HAP),
Initiative and Networking Fund of the Helmholtz Association,
Germany;
Fund for Scientific Research (FNRS-FWO),
FWO Odysseus programme,
Flanders Institute to encourage scientific and technological research in industry (IWT),
Belgian Federal Science Policy Office (Belspo);
Marsden Fund, New Zealand;
Australian Research Council;
Japan Society for Promotion of Science (JSPS);
the Swiss National Science Foundation (SNSF), Switzerland;
National Research Foundation of Korea (NRF);
Villum Fonden, Danish National Research Foundation (DNRF), Denmark

\newpage

% END: acknowledgments.tex

%\clearpage
%\bibliographystyle{apj}
%\bibliography{bib}

\begin{thebibliography}{}
\expandafter\ifx\csname natexlab\endcsname\relax\def\natexlab#1{#1}\fi

\bibitem[{Aartsen {et~al.}(2013{\natexlab{a}})}]{Aartsen:2013jdh}
Aartsen, M.~G., {et~al.} 2013{\natexlab{a}}, Science, 342, 1242856

\bibitem[{Aartsen {et~al.}(2013{\natexlab{b}})}]{Aartsen:2013rt}
---. 2013{\natexlab{b}}, Nucl. Instrum. Meth., A711, 73

\bibitem[{Aartsen {et~al.}(2014{\natexlab{a}})}]{Aartsen:2013vja}
---. 2014{\natexlab{a}}, JINST, 9, P03009

\bibitem[{Aartsen {et~al.}(2014{\natexlab{b}})}]{Aartsen:2014njl}
---. 2014{\natexlab{b}}, arXiv:1412.5106

\bibitem[{Aartsen {et~al.}(2014{\natexlab{c}})}]{Aartsen:2014gkd}
---. 2014{\natexlab{c}}, Phys. Rev. Lett., 113, 101101

\bibitem[{Aartsen {et~al.}(2014{\natexlab{d}})}]{Aartsen:2014cva}
---. 2014{\natexlab{d}}, Astrophys. J., 796, 109

\bibitem[{Aartsen {et~al.}(2015{\natexlab{a}})}]{hese4}
---. 2015{\natexlab{a}}, ICRC, 34, 1081

\bibitem[{Aartsen {et~al.}(2015{\natexlab{b}})}]{Aartsen:2014muf}
---. 2015{\natexlab{b}}, Phys. Rev., D91, 022001

\bibitem[{Aartsen {et~al.}(2015{\natexlab{c}})}]{Aartsen:2015rwa}
---. 2015{\natexlab{c}}, Phys. Rev. Lett., 115, 081102

\bibitem[{Aartsen {et~al.}(2015{\natexlab{d}})}]{Aartsen:2015ivb}
---. 2015{\natexlab{d}}, Phys. Rev. Lett., 114, 171102

\bibitem[{Aartsen {et~al.}(2015{\natexlab{e}})}]{Aartsen:2015wto}
---. 2015{\natexlab{e}}, Astrophys. J., 807, 46

\bibitem[{Aartsen {et~al.}(2016{\natexlab{a}})}]{Aartsen:2016qcr}
---. 2016{\natexlab{a}}, Astrophys. J., 824, 115

\bibitem[{Aartsen {et~al.}(2016{\natexlab{b}})}]{Aartsen:2016tpb}
---. 2016{\natexlab{b}}, Astrophys. J., 824, L28

\bibitem[{Aartsen {et~al.}(2016{\natexlab{c}})}]{Aartsen:2016xlq}
---. 2016{\natexlab{c}}, Astrophys. J., 833, 3

\bibitem[{Aartsen {et~al.}(2017{\natexlab{a}})}]{Aartsen:2016oji}
---. 2017{\natexlab{a}}, Astrophys. J., 835, 151

\bibitem[{Aartsen
  {et~al.}(2017{\natexlab{b}})}]{DETPAPER-1748-0221-12-03-P03012}
---. 2017{\natexlab{b}}, JINST, 12, P03012

\bibitem[{Abbasi {et~al.}(2009)}]{Abbasi:2008aa}
Abbasi, R., {et~al.} 2009, Nucl. Instrum. Meth., A601, 294

\bibitem[{Abbasi {et~al.}(2010)}]{Abbasi:2010vc}
---. 2010, Nucl. Instrum. Meth., A618, 139

\bibitem[{Abbasi {et~al.}(2012)}]{Collaboration:2011ym}
---. 2012, Astropart. Phys., 35, 615

\bibitem[{Achterberg {et~al.}(2006)}]{Achterberg:2006md}
Achterberg, A., {et~al.} 2006, Astropart. Phys., 26, 155

\bibitem[{Ackermann {et~al.}(2012)}]{Ackermann:2012:GammaProfile}
Ackermann, M., {et~al.} 2012, Astrophys. J., 750, 3

\bibitem[{Adrian-Martinez {et~al.}(2012)}]{AdrianMartinez:2012rp}
Adrian-Martinez, S., {et~al.} 2012, Astrophys. J., 760, 53

\bibitem[{Adrian-Martinez {et~al.}(2014)}]{Adrian-Martinez:2014wzf}
---. 2014, Astrophys. J., 786, L5

\bibitem[{Adrian-Martinez {et~al.}(2015)}]{TheANTARES:2015pba}
---. 2015, ICRC, 34, 1078

\bibitem[{Adrian-Martinez
  {et~al.}(2016{\natexlab{a}})}]{Adrian-Martinez:2016fdl}
---. 2016{\natexlab{a}}, J. Phys., G43, 084001

\bibitem[{Adrian-Martinez
  {et~al.}(2016{\natexlab{b}})}]{Adrian-Martinez:2015ver}
---. 2016{\natexlab{b}}, Astrophys. J., 823, 65

\bibitem[{Ahrens {et~al.}(2004)}]{Ahrens:2003fg}
Ahrens, J., {et~al.} 2004, Nucl. Instrum. Meth., A524, 169

\bibitem[{Becker(2008)}]{Becker:2007sv}
Becker, J.~K. 2008, Phys. Rept., 458, 173

\bibitem[{Chirkin \& Rhode(2004)}]{Chirkin:2004hz}
Chirkin, D., \& Rhode, W. 2004, arXiv:hep-ph/0407075

\bibitem[{Cooper-Sarkar {et~al.}(2011)Cooper-Sarkar, Mertsch, \&
  Sarkar}]{CooperSarkar:2011pa}
Cooper-Sarkar, A., Mertsch, P., \& Sarkar, S. 2011, JHEP, 08, 042

\bibitem[{Gaggero {et~al.}(2015)}]{Gaggero:2015xza}
Gaggero, D., {et~al.} 2015, Astrophys. J., 815, L25

\bibitem[{Gaisser {et~al.}(1995)Gaisser, Halzen, \&
  Stanev}]{Gaisser:Halzen:Stanev}
Gaisser, T.~K., Halzen, F., \& Stanev, T. 1995, Phys.Rept., 258, 173

\bibitem[{Glashow(1960)}]{Glashow:PhysRev.118.316}
Glashow, S.~L. 1960, Phys. Rev., 118, 316

\bibitem[{Learned \& Mannheim(2000)}]{Learned:Mannheim}
Learned, J., \& Mannheim, K. 2000, Ann. Rev. Nucl. Part. Sci., 50, 679

\bibitem[{Lehmann \& Romano(2005)}]{lehmann2005testing}
Lehmann, E.~L., \& Romano, J.~P. 2005, Testing statistical hypotheses, 3rd
  edn., Springer Texts in Statistics (New York: Springer), xiv+784

\bibitem[{{Neyman}(1937)}]{neyman}
{Neyman}, J. 1937, Philos. Tr. R. Soc. A, 236, 333

\bibitem[{Radel \& Wiebusch(2013)}]{Radel:2012ij}
Radel, L., \& Wiebusch, C. 2013, Astropart. Phys., 44, 102

\end{thebibliography}

\cleardoublepage

\begin{center}
  \begin{deluxetable*}{llrrrrr}
    \vspace{2em}
    \tablecaption{Summary of the source catalog search.  The objects are
      grouped by type, and within each type are sorted by increasing
      declination.   The type, common name, and equatorial coordinates
      (J2000) are shown for each object.  Where non-null ($\hat n_s>0$)
      results are found, the pre-trials significance $p_\text{pre}$ and
      best-fit $\hat n_s$ and $\hat\gamma$ are given.
    \vspace{.2em}\label{tab:cat}}
    \tablewidth{0pt}
    \tablehead{
      Type & Source & $\alpha~(^\circ)$ & $\delta~(^\circ)$
      & $p_\text{pre}$ & $\hat n_s$ & $\hat\gamma$
    }
    \startdata
    %\input{plots/cat.tex}
% cat.tex
BL Lac                     &PKS 2005-489        &    302.37&$    -48.82$&0.252   &   2.4&    2.2\\
                           &PKS 0537-441        &     84.71&$    -44.09$&0.256   &   1.7&    1.8\\
                           &PKS 0426-380        &     67.17&$    -37.93$&0.597   &   1.0&    1.8\\
                           &PKS 0548-322        &     87.67&$    -32.27$&0.634   &   1.2&    2.2\\
                           &H 2356-309          &    359.78&$    -30.63$&0.809   &   0.2&    2.4\\
                           &PKS 2155-304        &    329.72&$    -30.23$&0.642   &   1.2&    2.4\\
                           &1ES 1101-232        &    165.91&$    -23.49$&0.390   &   3.3&    2.8\\
                           &1ES 0347-121        &     57.35&$    -11.99$&0.543   &   2.5&    3.8\\
                           &PKS 0235+164        &     39.66&$     16.62$&$\cdot\cdot\cdot$&   0.0&$\cdot\cdot\cdot$\\
                           &1ES 0229+200        &     38.20&$     20.29$&$\cdot\cdot\cdot$&   0.0&$\cdot\cdot\cdot$\\
                           &W Comae             &    185.38&$     28.23$&0.618   &   0.6&    3.8\\
                           &Mrk 421             &    166.11&$     38.21$&$\cdot\cdot\cdot$&   0.0&$\cdot\cdot\cdot$\\
                           &Mrk 501             &    253.47&$     39.76$&0.404   &   1.5&    2.6\\
                           &\tablenotemark{$\dagger$}BL Lac              &    330.68&$     42.28$&0.010   &   6.9&    3.0\\
                           &H 1426+428          &    217.14&$     42.67$&0.566   &   0.5&    3.8\\
                           &3C66A               &     35.67&$     43.04$&0.482   &   0.9&    3.8\\
                           &1ES 2344+514        &    356.77&$     51.70$&0.189   &   2.9&    3.2\\
                           &1ES 1959+650        &    300.00&$     65.15$&0.519   &   0.6&    3.0\\
                           &S5 0716+71          &    110.47&$     71.34$&$\cdot\cdot\cdot$&   0.0&$\cdot\cdot\cdot$\\
\hline
Flat spectrum radio quasar &PKS 1454-354        &    224.36&$    -35.65$&0.612   &   1.6&    2.2\\
                           &PKS 1622-297        &    246.53&$    -29.86$&0.286   &   3.6&    2.2\\
                           &PKS 0454-234        &     74.27&$    -23.43$&$\cdot\cdot\cdot$&   0.0&$\cdot\cdot\cdot$\\
                           &QSO 1730-130        &    263.26&$    -13.08$&0.365   &   4.5&    3.8\\
                           &PKS 0727-11         &    112.58&$    -11.70$&$\cdot\cdot\cdot$&   0.0&$\cdot\cdot\cdot$\\
                           &PKS 1406-076        &    212.24&$     -7.87$&0.375   &   5.6&    3.8\\
                           &QSO 2022-077        &    306.42&$     -7.64$&$\cdot\cdot\cdot$&   0.0&$\cdot\cdot\cdot$\\
                           &HESS J1837-069      &    279.41&$     -6.95$&0.121   &   8.9&    3.8\\
                           &3C279               &    194.05&$     -5.79$&0.754   &   0.9&    3.8\\
                           &3C 273              &    187.28&$      2.05$&0.718   &   0.9&    2.8\\
                           &PKS 1502+106        &    226.10&$     10.49$&0.057   &   9.1&    3.8\\
                           &PKS 0528+134        &     82.73&$     13.53$&$\cdot\cdot\cdot$&   0.0&$\cdot\cdot\cdot$\\
                           &3C 454.3            &    343.49&$     16.15$&0.066   &   7.4&    3.8\\
                           &4C 38.41            &    248.81&$     38.13$&0.391   &   1.6&    2.4\\
\hline
Galactic center            &Sgr A*              &    266.42&$    -29.01$&0.080   &   5.6&    2.2\\
\hline
Not identified             &HESS J1507-622      &    226.72&$    -62.34$&0.473   &   0.7&    1.0\\
                           &HESS J1503-582      &    226.46&$    -58.74$&0.438   &   0.7&    1.0\\
                           &HESS J1741-302      &    265.25&$    -30.20$&0.072   &   5.7&    2.2\\
                           &HESS J1834-087      &    278.69&$     -8.76$&0.180   &   7.5&    3.8\\
                           &MGRO J1908+06       &    286.98&$      6.27$&0.078   &   8.5&    3.8\\
\hline
Pulsar wind nebula         &HESS J1356-645      &    209.00&$    -64.50$&0.795   &   0.1&    3.8\\
                           &PSR B1259-63        &    197.55&$    -63.52$&$\cdot\cdot\cdot$&   0.0&$\cdot\cdot\cdot$\\
                           &HESS J1303-631      &    195.74&$    -63.20$&$\cdot\cdot\cdot$&   0.0&$\cdot\cdot\cdot$\\
                           &MSH 15-52           &    228.53&$    -59.16$&0.408   &   0.7&    1.0\\
                           &HESS J1023-575      &    155.83&$    -57.76$&$\cdot\cdot\cdot$&   0.0&$\cdot\cdot\cdot$\\
                           &HESS J1616-508      &    243.78&$    -51.40$&0.166   &   2.4&    2.0\\
                           &HESS J1632-478      &    248.04&$    -47.82$&0.108   &   3.0&    2.0\\
                           &Vela X              &    128.75&$    -45.60$&$\cdot\cdot\cdot$&   0.0&$\cdot\cdot\cdot$\\
                           &Geminga             &     98.48&$     17.77$&$\cdot\cdot\cdot$&   0.0&$\cdot\cdot\cdot$\\
                           &Crab Nebula         &     83.63&$     22.01$&0.556   &   1.1&    2.8\\
                           &MGRO J2019+37       &    305.22&$     36.83$&0.224   &   3.5&    3.6\\
\hline
Star formation region      &Cyg OB2             &    308.08&$     41.51$&0.135   &   4.2&    3.4\\
\hline
Supernova remnant          &RCW 86              &    220.68&$    -62.48$&0.582   &   0.5&    1.0\\
                           &RX J0852.0-4622     &    133.00&$    -46.37$&$\cdot\cdot\cdot$&   0.0&$\cdot\cdot\cdot$\\
                           &RX J1713.7-3946     &    258.25&$    -39.75$&0.042   &   5.3&    2.2\\
                           &W28                 &    270.43&$    -23.34$&0.159   &   4.3&    2.2\\
                           &IC443               &     94.18&$     22.53$&$\cdot\cdot\cdot$&   0.0&$\cdot\cdot\cdot$\\
                           &Cas A               &    350.85&$     58.81$&0.261   &   2.0&    3.4\\
                           &TYCHO               &      6.36&$     64.18$&$\cdot\cdot\cdot$&   0.0&$\cdot\cdot\cdot$\\
\hline
Starburst/radio galaxy     &Cen A               &    201.37&$    -43.02$&0.629   &   1.0&    2.6\\
                           &M87                 &    187.71&$     12.39$&0.438   &   1.8&    2.6\\
                           &3C 123.0            &     69.27&$     29.67$&0.379   &   2.2&    3.0\\
                           &Cyg A               &    299.87&$     40.73$&0.276   &   2.6&    3.4\\
                           &NGC 1275            &     49.95&$     41.51$&0.479   &   1.0&    3.8\\
                           &M82                 &    148.97&$     69.68$&0.251   &   0.8&    2.0\\
\hline
Seyfert galaxy             &ESO 139-G12         &    264.41&$    -59.94$&0.096   &   3.0&    2.0\\
\hline
HMXB/mqso                  &Cir X-1             &    230.17&$    -57.17$&0.372   &   0.8&    1.0\\
                           &GX 339-4            &    255.70&$    -48.79$&0.052   &   4.3&    2.2\\
                           &LS 5039             &    276.56&$    -14.83$&0.444   &   1.7&    2.2\\
                           &SS433               &    287.96&$      4.98$&0.086   &   8.7&    3.8\\
                           &HESS J0632+057      &     98.25&$      5.80$&$\cdot\cdot\cdot$&   0.0&$\cdot\cdot\cdot$\\
                           &Cyg X-1             &    299.59&$     35.20$&0.382   &   2.2&    3.6\\
                           &Cyg X-3             &    308.11&$     40.96$&0.137   &   4.2&    3.4\\
                           &LSI 303             &     40.13&$     61.23$&$\cdot\cdot\cdot$&   0.0&$\cdot\cdot\cdot$\\
\hline
Massive star cluster       &HESS J1614-518      &     63.58&$    -51.82$&0.330   &   1.3&    1.6\\
% END: cat.tex
    \enddata
    \tablenotetext{\dagger}{Most significant source in the catalog, yielding
    $p_\text{post}=36\%$.}
  \end{deluxetable*}
\end{center}

\end{document}